\documentclass[letterpaper,11pt]{article}
\usepackage{amssymb, amsmath}
\usepackage{graphicx}
\usepackage{caption}
\usepackage{subcaption}
\usepackage{mathrsfs}
\usepackage{hyperref}
\usepackage{overcite}

\usepackage[letterpaper, portrait, margin=1.0in]{geometry}

\usepackage{color}

\def\f11{\mathstrut_1F_1}

\title{Generalization of the Ewens sampling formula to arbitrary fitness landscapes}

\author{Pavel Khromov$^{1,2}$, Constantin D. Malliaris$^{1,2}$ and Alexandre V. Morozov$^{1,2}$\footnote{Corresponding author: \texttt{morozov@physics.rutgers.edu}} \\
        \small{$^1$ Department of Physics and Astronomy, Rutgers University, Piscataway, NJ 08854, USA} \\
        \small{$^2$ Center for Quantitative Biology, Rutgers University, Piscataway, NJ 08854, USA}
        }

\date{}

\begin{document}

\maketitle

\begin{abstract}
In considering evolution of transcribed regions, regulatory modules, and other
genomic loci of interest, we are often faced with a situation in which 
the number of allelic states greatly exceeds the population size. In this limit,
the population eventually adopts a steady state characterized by mutation-selection-drift balance. Although new alleles continue to be explored through mutation,
the statistics of the population, and in particular the probabilities of seeing
specific allelic configurations in samples taken from a population,
do not change with time. In the absence of selection, probabilities
of allelic configurations are given by the Ewens sampling formula, widely used in population genetics to detect deviations from neutrality. Here we
develop an extension of this formula to arbitrary, possibly epistatic, fitness landscapes.
Although our approach is general, we focus on the class of landscapes
in which alleles are grouped into two, three, or several fitness states.
This class of landscapes yields sampling probabilities that are computationally more tractable, and can form a basis for the inference of selection signatures from sequence
data.
We demonstrate that, for a sizeable range of mutation rates and selection coefficients,
the steady-state allelic diversity is not neutral. Therefore, it may be used
to infer selection coefficients, as well as other key evolutionary parameters,
using high-throughput sequencing of evolving populations to collect data on
locus polymorphisms. We also carry out numerical investigation of
various approximations involved in deriving our sampling formulas, such as
the infinite allele limit and the ``full connectivity'' assumption in which
each allele can mutate into any other allele. We find that
our theory remains sufficiently accurate even if these assumptions are relaxed.
Thus, our framework establishes a theoretical foundation for inferring selection 
signatures from samples of sequences produced by evolution on epistatic fitness
landscapes.
\end{abstract}

\section*{Introduction}

With the advent of high-throughput molecular biology techniques, it has recently
become possible to carry out large-scale phenotypic assays in molecular systems. For example, Podgornaia and Laub have mapped all $20^4 = 1.6 \times 10^5$
possible combinations of four key residues in the \textit{E. coli} protein kinase PhoQ, and assayed
each mutant for the signaling function mediated by its binding partner PhoP~\cite{Podgornaia2015}. The study revealed $1659$ functional PhoQ variants,
which can be thought of as forming the upper plane on the fitness landscape. The upper plane is divided into
several clusters under single-point mutational moves -- only sequences within each cluster can mutate into each other without undergoing deleterious moves to the lower plane,
where all non-functional sequences reside. This two-plane landscape is highly epistatic --
the effect of a given mutation depends on the amino acids at the other three positions, in agreement with previous reports on the primary role of epistasis in
molecular evolution~\cite{Lunzer2005,Romero2009,Lunzer2010,Breen2012}.

The picture of fitness landscapes made of multiple interconnected planes is supported by
other high-throughput experiments aimed at elucidation of the relationship between gene sequence and function~\cite{Romero2009,Lind2010,Hietpas2011}. 
Although these experiments typically yield distributions of mutation fitness effects,
the data can often be understood, at least to a first-order approximation,
in terms of functional and non-functional sequence variants~\cite{Sanjuan2004,EyreWalker2007}. Indeed, distributions of fitness effects
often appear to be bi-modal in such studies, with a low-fitness peak for strongly
deleterious and lethal mutations, and another for weakly deleterious and neutral
ones. In comparison to deleterious and neutral mutations, beneficial mutations are
relatively infrequent~\cite{Lind2010,Hietpas2011,EyreWalker2007}.
The coarse-graining of the fitness landscape into functional and non-functional states
may be refined by introducing additional fitness values, e.g. for weakly deleterious
mutations.

Overall, given the astronomically large number of possible sequence variants,
we expect the size of neutrally-connected clusters on all fitness planes
to be larger than the population size. Then evolutionary dynamics on a multiple-plane landscape will involve
periods of neutral search followed by positive selection, in which the bulk of the population moves to a higher-fitness state~\cite{Romero2009,Wagner2008}.
Populations evolving on such a landscape will eventually reach a steady state characterized by mutation-drift balance~\cite{ewens-04}. This balance determines
the statistics of the population, such as the mean and the variance of the number
of distinct alleles. Although the population continues to explore new alleles
through mutations, the allele statistics do not change anymore once the steady state
is reached. In the absence of selection (that is, for evolution on a single neutral plane), the probability of observing a given pattern of allelic diversity in a sample of size $n$ taken from the steady-state population was derived by Ewens~\cite{ewens-72}.

The Ewens sampling formula can be used to understand the allelic diversity in
neutral populations, and test for deviations from the neutral expectation~\cite{Slatkin1994}. However, in order to make quantitative predictions
of selection coefficients, it is necessary to extend the Ewens sampling formula
to arbitrary fitness landscapes. As noted above, of special interest in molecular
evolution are landscapes in which alleles are grouped into two or three distinct fitness states. Such landscapes provide a natural generalization of the completely neutral evolutionary scenario.

Previous work in this area has focused mostly on deriving frequency spectra for either
arbitrary fitness landscapes or specific models of selection. In particular,
Li obtained the frequency spectrum for a general landscape,
and used it to derive expressions for the mean number of alleles in a sample,
as well as the mean and the variance of the heterozygosity~\cite{li-77,li-78,li-79}.
Ewens and Li derived frequency spectra for landscapes with two
and three fitness states, and used them to compute the mean number of distinct alleles
and the mean heterozygosity~\cite{ewens-80}.
Griffiths also derived a general integral expression for the frequency
spectrum~\cite{griffiths-83}.

More recently, Ethier and Kurtz have studied allelic diversity in
a general model of selection in which fitness of each new allele is completely independent of the fitness of its parent.~\cite{ethier-87} This and follow-up
work~\cite{joyce-95a,joyce-95b,grote-02} has contributed to our understanding of how
allelic sampling probabilities are shaped by various forms of selection pressure.
In particular, Joyce and Genz have developed effective algorithms
for evaluating sampling probabilities~\cite{joyce-05}.
Finally, Desai et al. have investigated sampling probabilities in a model
(previously introduced by Charlesworth et al.~\cite{charlesworth-93} and Hudson and Kaplan~\cite{Hudson1994}) comprised of a sequence of neutral and negatively selected sites~\cite{desai-12}. This model has no epistasis, and therefore can be treated using
the Poisson Random Field method~\cite{sawyer-92}. However, the models of both Ethier and Kurtz and Desai et al. cannot be applied to molecular evolution, which is
characterized by prominent epistasis and correlated fitness values.

Here we develop an extension of the Ewens sampling formula to arbitrary fitness
landscapes. First, we use the diffusion approximation of population dynamics
to derive a general sampling formula valid for
any number of alleles $K$ (allele $i$ is assigned an arbitrary fitness value $f_i$), mutation rate $\mu$, and population size $N$.
We assume that the population adopts a steady state characterized by mutation-selection-drift balance.
The sampling formula is derived under
the assumption that $n \ll N$, where $n$ is the total sample size; this assumption is
often realistic in populations subjected to high-throughput sequencing.

The most general sampling formula is not amenable to efficient calculations
since it involves sums over special functions with a number of terms that
increases rapidly with both the sample size and the number of alleles. Therefore, we focus on multiple-plane
landscapes applicable in molecular evolution, with only a few (two or more) distinct fitness states. We also make the $N \ll K$ approximation, which
reflects the fact that the number of possible allelic states in molecular systems
is typically much larger than effective population sizes.
The resulting sampling
formulas are sufficiently tractable to be used to study selection signatures
and deviations from neutrality on multiple-plane fitness landscapes for arbitrary
mutation rates and selection coefficients. In particular, we study the effective population size approximation~\cite{charlesworth-93,desai-12} and its limits of
applicability. We compare our results with numerical simulations, investigating
potential deviations between experiment and theory which may be caused by
the differences in evolutionary dynamics between fully connected sequence networks (for which our theory is valid) and more realistic scenarios involving single-point mutations.
We also investigate finite network size effects, since our multiple-plane
sampling formula is derived in the infinite allele limit.

Our results are applicable to understanding the nature of allelic diversity
under selection, mutation and drift, for a vast class of fitness landscapes
that are relevant to both molecular evolution and, more generally, evolution in
systems where the number of alleles vastly exceeds the population size.
Moreover, our sampling formulas form the basis for a quantitative test
which can both detect the presence of selection and estimate selection
coefficients in epistatic systems under very general and well-defined assumptions. 
Population-level allele diversity data is increasingly available through
high-throughput sequencing techniques, making our approach a practical and timely tool
for studying the role played by selection in present-day populations.

\section*{Results}

\subsection*{Steady-state distribution of allele frequencies}

We consider a haploid population of fixed size $N$. Each individual in the population
has a single allele in the state $i$, with fitness $f_i$; there are $K$ distinct allelic states.
Mutations occur at a probability $\mu$ per generation, replacing the original allele with
one of the $K-1$ remaining alleles. Thus the probability of offspring $A_j$
produced by parent $A_{i \ne j}$ is ${\mu}/{(K-1)}$. We can view this system as an ``allelic network''
with the topology of a complete graph,
with $K$ vertices representing allelic states and edges representing mutational moves.
Stochastic evolution of the population can then be described using Moran ~\cite{moran-58,Gillespie2004} or Wright-Fisher~\cite{Gillespie2004,wright-31} population dynamics.

Without loss of generality, we can specify fitness $f_i$ of the allele $A_i$
with respect to an arbitrary reference allele $A_K$. It is convenient to
introduce a $K$-dimensional vector of relative fitnesses multiplied by the population size:
$\vec\beta = (N(f_1-f_K), ..., N(f_{K-1}-f_K), 0)$. Likewise, we define a $K$-dimensional
vector of mutation rates as $\vec\epsilon = (\epsilon, ..., \epsilon)$,
where $\epsilon = {N \mu}/{(K-1)}$ for Moran~\cite{moran-58,Gillespie2004} and
$\epsilon = {2N \mu}/{(K-1)}$ for Wright-Fisher population dynamics~\cite{Gillespie2004,wright-31}.
\footnote{All the formulas in this section can be generalized to the case of final-state-dependent mutation rates, i.e. $\mu_{ij} = \mu_j$, $\forall i$ in $A_i \to A_j$}
We also introduce $\theta = K \epsilon$, which
in the limit of large number of alleles $K \to \infty$ becomes
$N \mu$ and $2 N \mu$ for Moran and Wright-Fisher dynamics, respectively.
Note that we consider the case of equal mutation rates between alleles, for which
the steady state is well-defined~\cite{lassig-10}. 

The evolutionary dynamics of this system in the diffusion limit is described by the
forward Kolmogorov equation: 
\begin{equation}\label{fokker-plank}
	\frac{\partial p}{\partial t}
	=
	\frac 12 
	\sum_{i,j=1}^{K-1}
	\frac{\partial^2 (V_{ij} p)}{\partial x_i \partial x_j}
	-\sum_{i=1}^{K-1}
	\frac{\partial (M_{i} p)}{\partial x_i},
\end{equation}
where $p(\vec{x}, t)$ is the joint probability of frequencies of $K$ alleles at time $t$ ($\vec{x} = (x_1, ..., x_K)$ is a $K$-dimensional vector of allele frequencies which occupy a $(K-1)$-dimensional simplex
$\Sigma_{K-1} = \{ (x_1\ge 0, ..., x_K\ge 0) : \sum_{i=1}^K x_i = 1 \}$), and

\begin{equation}
\begin{split}
	M_i &= \mathbb{E}[\delta x_i] =
	\frac{\epsilon - K\epsilon x_i}{2N}
	+\frac 12 x_i
	\left[
	\frac{\partial \langle f \rangle}{\partial x_i}
	-
	\sum_{j=1}^{K-1} x_j \frac{\partial \langle f \rangle}{\partial x_j}
	\right]
	+
	O \left( \frac{1}{N^2} \right), \\
	V_{ii} &= \mathbb{E}[\delta x_i^2] - \mathbb{E}[\delta x_i]^2 =
	\frac{x_i (1-x_i)}{N}
	+
	O \left( \frac{1}{N^2} \right), \\
	V_{ij} &= \mathbb{E}[\delta x_i \delta x_j] - \mathbb{E}[\delta x_i] \mathbb{E}[\delta x_j] =
	-\frac{x_i x_j}{N}
	+
	O \left( \frac{1}{N^2} \right), \\
\end{split}
\end{equation}
where $x_i$ denotes the frequency of allele $A_i$ in the population, and
$\langle f\rangle = \sum_{i=1}^K f_i x_i $
is the mean population fitness.

In steady state $\partial p / \partial t = 0$, and the
distribution of allele frequencies in Eq.~\ref{fokker-plank} is given
by~\cite{li-77,li-78,watterson-77}
\begin{equation}\label{watterson-distribution}
  p(\vec{x})
  =
  \frac{1}{Z}
  e^{N \langle f \rangle}
  \prod_{i=1}^K x_i^{\epsilon-1},
\end{equation}
where $Z$ is the normalization constant.
Eq.~\ref{watterson-distribution} can be written more explicitly as
\begin{equation}\label{allele-frequency-distribution}
  p(\vec{x}) =
  \frac{1}{B(\vec\epsilon) \mathcal F(\vec\epsilon; |\vec\epsilon|; \vec\beta)}
  \prod_{i=1}^K x_i^{\epsilon - 1} e^{\beta_i x_i},
\end{equation}
where $|\vec\epsilon| = K \epsilon = \theta$ is the 
$L_1$-norm of $\vec\epsilon$,
\begin{equation}
  B(\vec a) = \frac{\prod_{i=1}^n \Gamma(a_i)}{\Gamma(\sum_{i=1}^n a_i)}
\end{equation}
is the generalized beta function written in terms of Gamma functions, and
\begin{equation} \label{gen-hyp-def}
 \mathcal F(\vec a; b; \vec z)
 =
 \sum_{j_1=0}^\infty ... \sum_{j_n=0}^\infty
 \frac{a_1^{(j_1)}...a_n^{(j_n)}}{b^{(j_1+...+j_n)}}
 \frac{z_1^{j_1}}{j_1!} ... 
 \frac{z_n^{j_n}}{j_n!} =
 \sum_{j=0}^\infty \frac{B_j(\alpha_1, ..., \alpha_j)}{j! b^{(j)}}
\end{equation}
is the confluent hypergeometric function $\f11(a;b;z)$ generalized to
vector arguments. Here, $a^{(j)}=\Gamma(a+j)/\Gamma(a)$ is the rising factorial,
$B_j$ is the $j$th complete Bell polynomial and $\alpha_j = (j-1)! \sum_{i=1}^n a_i z_i^j$.

Differentiation of this function with respect to $\vec z$ yields
\begin{equation} \label{calF:deriv}
	\prod_{i=1}^k
	\left(\frac{\partial}{\partial z_i} \right)^{n_i}
	\mathcal F(\vec a; b; \vec z)
	=
	\frac{\prod_{i=1}^k (a_i)^{(n_i)}}{b^{(n)}}
	\mathcal F \left(\vec a + \sum_{i=1}^k n_i \vec 1_i; b+n; \vec z \right),
\end{equation}
where $n=\sum_{i=1}^k n_i$ and $(1_i)_j = \delta_{ij}$.

\subsection*{Strongly monomorphic limit}

When mutation rate decreases and population size is kept fixed, $\epsilon \to 0$ and the population becomes monomorphic~\cite{Gillespie2004,kimura-62,kimura-69,Crow1970}. We consider the 
Fourier transform of the steady-state distribution in Eq.~\ref{allele-frequency-distribution}:
\begin{equation}
  \tilde p(\vec k) = \int_{\Sigma_{K-1}} d\vec x e^{i \vec k \cdot \vec x} p(\vec x),
\end{equation}
where the integral is over the $(K-1)$-dimensional simplex. Using Eq.~\ref{gen-hyp-def}, we can write the 
Fourier transform as a ratio of two generalized hypergeometric functions:
\begin{equation} \label{eq:Fratio}
  \tilde p(\vec k) = 
  \frac
  {\mathcal F(\vec\epsilon; |\vec\epsilon|; \vec\beta+i\vec k)}
  {\mathcal F(\vec\epsilon; |\vec\epsilon|; \vec\beta)}.
\end{equation}

Taking the $\epsilon \to 0$ limit yields
\begin{equation}
  \tilde p_{\mathrm{mono}} (\vec k) = 
  \frac
  {\sum_{m=1}^K e^{\beta_m+i k_m}}
  {\sum_{m=1}^K e^{\beta_m}}.
\end{equation}
Thus the steady-state distribution
in the monomorphic limit is given by:
\begin{equation} \label{strongly-monomorphic-limit}
  p_{\mathrm{mono}}(\vec x) = \int \frac{d\vec x}{\mathrm{Vol}(\Sigma_{K-1})}
  e^{-i \vec k \cdot \vec x} \tilde p_{\mathrm{mono}} (\vec k)
  = \frac
  {\sum_{m=1}^K e^{\beta_m} \delta(\vec x-\vec 1_m)}
  {\sum_{m=1}^K e^{\beta_m}},
\end{equation}
where $\mathrm{Vol}(\Sigma_{K-1}) = \sqrt{K}/(K-1)!$ is the volume of the $(K-1)$-dimensional simplex
and $(1_m)_i = \delta_{mi}$. Therefore in the $\epsilon \to 0$ limit the population resides in one of
the $K$ monomorphic states available to it, with the probability of being in a particular
state exponentially weighted by its fitness~\cite{sella-05,sella-09,rouzine-01}. 

\subsection*{Probability of a sample of alleles}

Let us now consider a situation relevant to molecular evolution, where the number of alleles
$K$ is much larger than the population size $N$.
In this case, the steady state in terms of allele frequencies
is unlikely to be reached on relevant evolutionary time scales.
Mathematically, the $K \to \infty$ limit of
Eq.~\ref{allele-frequency-distribution} becomes ill-defined~\cite{kingman-75,kingman-77}.
Nonetheless, the steady state is well-defined in terms of allelic
\emph{counts} rather than frequencies of specific alleles~\cite{ewens-04}.
In other words, the allelic diversity of the population (e.g. as characterized
by the mean and the variance of the distribution of the number of distinct allelic types) is
tractable and will no longer change in steady state, although new
alleles will continue being explored through mutation.

One is often interested in statistical properties of a sample of alleles
of size $n \ll N$ obtained from the population.
Let us consider a simple example of a population
evolving on a small allelic network with $K=5$ allelic types $A,B,C,D,E$.
Suppose that in sampling $n=4$ alleles from the population we first observe allele $A$, then $C$, then $A$ again, and finally $D$.
We can record this sequence of alleles as an ordered list $(A,C,A,D)$.
However, typically we are not interested in the order in which alleles
appear in the sample, and therefore record the result as an unordered list $\{A,A,C,D\}$,
which shows that allele $A$ has appeared twice and alleles $C$ and $D$ have appeared once each.
\footnote{We shall use the notation $\{a,b,...,z\}$ for unordered lists and
$(a,b,...,z)$ for ordered ones.
For ordered lists $(a,b,...,z) \ne (b,a,...,z)$, whereas
for unordered lists $\{a,b,...,z\} = \{b,a,...,z\}$.}
Alternatively, we can record non-zero allelic counts,
which gives us $n_A=2,n_C=1,n_D=1$. Finally, we can dispense with the allele labels
altogether, identifying each allele in the sample as either new or already seen.
In this case, we are left with an unordered list of counts $\{2,1,1\}$,
meaning that we have observed
$4$ alleles of $3$ different types, with one type represented by two alleles and
the other two types by one each. In general, we will refer to
$\{n_1, ..., n_k\}$ as the (unordered) allelic counts.
The allelic counts can also be summarized in terms of a histogram which records
how many groups of $j$ identical alleles occur in the sample,
with $j$ ranging from $1$ to $n$.
In our example, there is one group of two identical alleles and
two groups of one allele each, so that
$(A,C,A,D)$ is recorded as the allelic histogram $(a_1=2, a_2=1, a_3=0, a_4=0)$.
\footnote{We shall use the notation $(a_1,a_2,...,a_n)$ for allelic histograms.}


We now derive the probability $\mathbb{P} [\{n_1, ..., n_k\}]$ of observing an unordered sample $\{n_1, ..., n_k\}$, given that the population has reached steady state in terms
of its allelic diversity. Before treating the general case, we illustrate our approach using a toy example with $K=3$ allelic types: $\mathcal{A} = (A,B,C)$. We wish to calculate the probability of observing the $\{2,1\}$ unordered configuration in a
sample of size $n=3$, which is assumed to be much less than the population size $N$.
There are $18$ ordered configurations that contribute to this probability:
\begin{center}
  \begin{tabular}{c c c}
    $AAB$&$ABA$&$BAA$\\
    $AAC$&$ACA$&$CAA$\\
    $BBC$&$BCB$&$CBB$\\
    \\
    $ABB$&$BAB$&$BBA$\\
    $ACC$&$CAC$&$CCA$\\
    $BCC$&$CBC$&$CCB$
  \end{tabular}
\end{center}
In particular, the probability of choosing $A$ first, then $A$ again and finally $B$ is
\begin{equation}
  \begin{split}
    \mathbb P[(A,A,B)] &= \int x_A^2 x_B^1 \; p(x_A, x_B, x_C) \; dx_A dx_B dx_C \\
    &= \int x_A^2 x_B^1 \; p(x_A, x_B) \; dx_A dx_B,
  \end{split}
\end{equation}
where $p(x_A, x_B, x_C)$ is given by Eq.~\ref{allele-frequency-distribution}.
Consequently, the probability of observing two $A$'s and one $B$ in \emph{any} order is
given by~\cite{etheridge-11}
\begin{equation}
	\mathbb P[\{A,A,B\}] = \binom{3}{2 \; 1} \int x_A^2 x_B^1 \; p(x_A, x_B) \; dx_A dx_B,
\end{equation}
where $\binom{3}{2 \; 1}$ is the multinomial coefficient.
Introducing a set $S_2\mathcal A=\{ (A,B), (A,C), (B,C)\}$, which permutes allelic
identities in an ordered manner (i.e., the overall allele ordering from $A$ to $B$ to $C$ is preserved in each pair of alleles), we can take into account the first $9$ configurations in the table above:
\begin{equation} \label{half:prob}
	\mathbb P[\{A,A,B\}] + \mathbb P[\{A,A,C\}] + \mathbb P[\{B,B,C\}] = 
    \binom{3}{2 \; 1}
    \sum_{\sigma \in S_2\mathcal A}\int x_{\sigma_1}^2 x_{\sigma_2}^1 \;
    p(x_{\sigma_1}, x_{\sigma_2}) \; dx_{\sigma_1} dx_{\sigma_2}.
\end{equation}
In order to include the last $9$ configurations in the table, we need to switch
the order of the alleles: $\{ (A,B), (A,C), (B,C)\} \to \{ (B,A), (C,A), (C,B)\}$.
But switching the alleles in each pair amounts to replacing
$x_{\sigma_1}^2 x_{\sigma_2}^1$ with $x_{\sigma_2}^2 x_{\sigma_1}^1 = x_{\sigma_1}^1 x_{\sigma_2}^2$ in Eq.~\ref{half:prob}. Thus we can summarize the entire table
by introducing a set $P\mathcal{N}$ of all distinct permutations of allelic counts $\mathcal N$, which determine the powers to which
the allelic frequencies are raised in Eq.~\ref{half:prob}. In our example
$\mathcal N = (2,1)$ and $P\mathcal{N} \equiv \{\nu_1,\nu_2\} = \{(2,1), (1,2)\}$. Therefore,
\begin{eqnarray}
	\mathbb P[\{2,1\}] &=&
    \binom{3}{2 \; 1}
    \sum_{\nu \in P\mathcal N}
    \sum_{\sigma \in S_2\mathcal A}\int x_{\sigma_1}^{\nu_1} x_{\sigma_2}^{\nu_2} \;
    p(x_{\sigma_1}, x_{\sigma_2}) \; dx_{\sigma_1} dx_{\sigma_2} \\  
    &=& \binom{3}{2 \; 1}
	\sum_{\nu\in P\mathcal N}
    \sum_{\sigma\in S_2 \mathcal A}
    \mathbb{E}\left[ \prod_{i=1}^2 x_{\sigma_i}^{\nu_i} \right].
\end{eqnarray}

The above example can be easily generalized to describe the probability
$\mathbb P[\{n_1, ..., n_k\}]$ of observing an unordered list of counts, $\{n_1, ..., n_k\}$. Note that the sample size is $n=\sum_{i=1}^k n_i$, and that $k$ distinct allelic types
are observed. First, we enumerate all $K$ alleles, forming a unique
ordered list $\mathcal A = (1, ... ,K)$. Second,
we choose a subset $\sigma = (\sigma_1, ..., \sigma_k)$
of size $k$ from $\mathcal A$ without replacement,
so that the allelic order is preserved: $\sigma_1 < ... < \sigma_k$
(note that no subsets are allowed to contain repeating elements of $\mathcal A$).
Then $S_k \mathcal{A}$ can be naturally defined as a set which contains
all ordered subsets of $\mathcal A$ of size $k$. Finally, as before
$P\mathcal{N} \equiv \{\nu_1, ...,\nu_k\}$ is a set of all distinct permutations of 
allelic counts $\mathcal N = (n_1, ..., n_k)$. With these definitions,
\begin{equation} \label{Pn}
	\mathbb P [\{n_1, ..., n_k\}] = 
	\binom{n}{n_1 \; ... \; n_k}
	\sum_{\nu\in P\mathcal N}
    \sum_{\sigma\in S_k \mathcal A}
    \mathbb{E}\left[ \prod_{i=1}^k x_{\sigma_i}^{\nu_i} \right],
\end{equation}
where $\binom{n}{n_1 \; ... \; n_k}$ is the multinomial coefficient, and the expectation is calculated with respect to the steady-state allele distribution,
Eq.~\ref{allele-frequency-distribution}.

We can use the probability distribution over unordered configurations (Eq.~\ref{Pn}) to
compute the distribution of the number of different allelic types $k$:
\begin{equation}
	\mathbb P[k]
	=
	\sum_{\substack{n_1 \ge ... \ge n_k \\ n_1+...+n_k=n}}
	\mathbb P[\{n_1, ..., n_k\}],
\end{equation}
where the summation runs over
all ordered partitions of $n$ into $k$ positive integers.

\subsection*{Sampling formula for the arbitrary fitness landscape}

As Eq.~\ref{Pn} demonstrates, evaluation of sample probabilities requires
calculation of moments of allele frequency distributions. This is most easily accomplished by
taking derivatives of the normalization constant
$Z = B(\vec\epsilon) \mathcal{F}(\vec\epsilon; |\vec\epsilon|; \vec\beta)$
with respect to the components of $\vec\beta$:
\begin{equation}
	\mathbb{E}\left[\prod_{i=1}^k x_i^{\nu_i}\right]
	=
	\frac 1Z
	\prod_{i=1}^k
	\left( \frac{\partial}{\partial \beta_i} \right)^{\nu_i}
	Z.
\end{equation}
Using Eq.~\ref{calF:deriv},
we obtain:
\begin{equation}\label{sampling-prob}
\begin{split}
	\mathbb{P}[\{n_1, ..., n_k\}] =
	\binom{n}{n_1 \; ... \; n_k}
	\frac{\prod_{i=1}^k \epsilon^{(n_i)}}{(K\epsilon)^{(n)}}
	\sum_{\nu\in P\mathcal N}
	\sum_{\sigma\in S_k \mathcal A}
	\frac{\mathcal F(\vec\epsilon + \vec \nu_\sigma; K\epsilon + n; \vec\beta)}
	{\mathcal F(\vec\epsilon; K\epsilon; \vec\beta)},
\end{split}
\end{equation}
where equal mutation rates are assumed for all alleles and 
$\vec\nu_\sigma$ is a $K$-dimensional vector whose $\sigma_i$-th
components are $\nu_i$ ($i=1,...,k$) and all the other components are zero. As discussed above,
the sum over $\sigma$ extends over all distinct subsets of $k$ alleles
sampled from $K$ uniquely ordered alleles and
subject to the $\sigma_1 < ... < \sigma_k$ constraint.
Therefore $\vec\nu_\sigma$ has $K-k$ zero and $k$ non-zero components
which are distributed according to $\sigma$.
The sum over $\nu$ extends over all distinct permutations 
of allelic counts which sum up to $n$. Eq.~\ref{sampling-prob} is valid
for arbitrary fitness landscapes and arbitrary $K$.

\subsection*{Neutral limit of the sampling formula}

When all alleles have the same fitness, the 
general sampling formula given by Eq.~\ref{sampling-prob} should reduce to the
Ewens formula for neutral evolutionary dynamics~\cite{ewens-04,ewens-72}.
Indeed, with all $\beta_i$ set to zero, the generalized hypergeometric function
$\mathcal F(\vec a; b; \vec 0)$ (Eq.~\ref{gen-hyp-def}) becomes $1$.
Then for the finite number of alleles $K$

\begin{equation} \label{sampling-prob-neutral-finite-K}
  \mathbb{P}[\{n_1, ..., n_k\}] =
  N_P
  \frac{n!}{(K\epsilon)^{(n)}}
  \binom{K}{k}
  \prod_{i=1}^k \frac{\epsilon^{(n_i)}}{n_i!},
\end{equation}
where $N_P = |P\mathcal N|$ is the total number of distinct permutations of allelic counts
$(n_1, ..., n_k)$. In the limit of an infinite number of alleles $K\to\infty$,
Eq.~\ref{sampling-prob-neutral-finite-K} becomes
\begin{equation} \label{sampling-prob-neutral}
  \mathbb{P}[\{n_1, ..., n_k\}]
  =
  N_P
  \frac{1}{k!}
  \frac{n!}{\prod_{i=1}^k n_i}
  \frac{\theta^k}{\theta^{(n)}}.
\end{equation}
Changing variables to allelic histogram counts yields
$\prod_{i=1}^k n_i = \prod_{j=1}^n j^{a_j}$
and
$N_P = k!/\prod_{j=1}^n a_j!$,
resulting in
\begin{equation} \label{sampling-prob-neutral-hist}
  \mathbb{P}[(a_1, ..., a_n)]
  =
  \frac{n!}{\prod_{j=1}^n a_j! j^{a_j}}
  \frac{\theta^k}{\theta^{(n)}}.
\end{equation}
Eq.~\ref{sampling-prob-neutral-hist} is known as the Ewens sampling formula
\cite{ewens-04,ewens-72}.


%


\subsection*{Sampling formula for populations with two fitness states} 

As a straightforward generalization of the neutral case,
consider a system with $I$ alleles of fitness $f_1$ and 
$K-I$ alleles with fitness $f_2 \ne f_1$. Thus the fitness
landscape consists of two interconnected ``planes''. We can assume without loss of generality that alleles $1, \dots ,I$ belong to the first plane
and alleles $I+1, \dots ,K$ belong to the second plane.
Then $\gamma = I/K$ defines a fraction of nodes on the first plane, and
the fitness vector is $\vec\beta = (\beta, ..., \beta, 0, ..., 0)$
with $I$ non-zero entries followed by $K-I$ zeros. Then it can be shown
that for finite $K$ the sampling probability is given by 
(Appendix~\ref{appendix:sampling-two-planes}):

\begin{equation} \label{sampling-prob-2-planes-finite-K}
\begin{split}
 \mathbb{P}[\{n_1, ..., n_k\}&]
 =
 \frac{n!}{\prod_{i=1}^k n_i!}
 \frac{\prod_{i=1}^k \epsilon^{(n_i)}}{(K\epsilon)^{(n)}}
 \binom{K}{k} \times
 \\
 &
 \sum_{\nu\in P\mathcal N}
 \sum_{i=0}^k
 \frac{\f11 \left( \gamma\theta+\sum_{m=1}^i \nu_m; \theta+n; \beta \right)}
 {\f11\left(\gamma \theta; \theta; \beta \right)}
 \frac{\binom{I}{i} \binom{K-I}{k-i}}{\binom{K}{k}}.
\end{split}
\end{equation}

Taking the infinite allele ($K \to \infty$) limit with $\gamma$ fixed, we arrive at
\begin{equation} \label{sampling-prob-2-planes}
\begin{split}
 \mathbb{P}[\{n_1, ..., n_k\}&]
 =
 \frac{n!}{k!}
 \frac{1}{\prod_{i=1}^k n_i}
 \frac{\theta^k}{\theta^{(n)}} \times
 \\
 &
 \sum_{\nu\in P\mathcal N}
 \sum_{i=0}^k
 \frac{\f11 \left( \gamma\theta+\sum_{m=1}^i \nu_m; \theta+n; \beta \right)}
 {\f11\left(\gamma \theta; \theta; \beta \right)}
 \binom{k}{i}
 \gamma^i (1-\gamma)^{k-i}.
\end{split}
\end{equation}
Thus hypergeometric sampling of Eq.~\ref{sampling-prob-2-planes-finite-K}
reduces to binomial sampling in the infinite allele limit.

\subsection*{Sampling formula for populations with multiple fitness states}

Let us now generalize the result of the previous section to
the case of multiple fitness states: each allele can be assigned a distinct fitness value $f_m$,
$m = 1, \dots ,M$. In other words, the fitness landscape consists of
multiple planes, with $I_m = \gamma_m K$ nodes of fitness $f_m$ on the $m$th
plane, so that $\sum_{m=1}^M \gamma_m = 1$.
Then the sampling probability for finite $K$ is given by
\begin{equation} \label{sampling-prob-mult-planes-finite-K}
\begin{split}
  \mathbb P[\{n_1, ..., n_k\}&]
  =
  \frac{n!}{\prod_{i=1}^k n_i!}
  \frac{\prod_{i=1}^k \epsilon^{(n_i)}}{(K\epsilon)^{(n)}}
  \binom{K}{k} \times \\
  &
  \sum_{\nu\in P\mathcal N}
  \sum_{Y\in \mathcal Y(\vec I, \mathcal N)}
  \frac{\mathcal F(\vec\gamma \theta + \vec \nu^Y;\theta+n; \vec\beta)}
  {\mathcal F (\vec\gamma \theta; \theta; \vec\beta)}
  \frac{\binom{I_1}{i_1} ... \binom{I_M}{i_M}}{\binom{K}{k}},
\end{split}
\end{equation}
and its infinite allele limit is given by
\begin{equation} \label{sampling-prob-mult-planes}
\begin{split}
  \mathbb P[\{n_1, ..., n_k\}&]
  =
  \frac{n!}{k!}
  \frac{1}{\prod_{i=1}^k n_i}
  \frac{\theta^k}{\theta^{(n)}} \times \\
  &
  \sum_{\nu\in P\mathcal N}
  \sum_{Y\in \mathcal Y(\mathcal N)}
  \frac{\mathcal F(\vec\gamma \theta + \vec \nu^Y;\theta+n; \vec\beta)}
  {\mathcal F (\vec\gamma \theta; \theta; \vec\beta)}
  \binom{k}{i_1 ... i_M}
  \gamma_1^{i_1} ... \gamma_M^{i_M}.
\end{split}
\end{equation}



\begin{figure}
        \centering
        	\includegraphics[width=\textwidth]{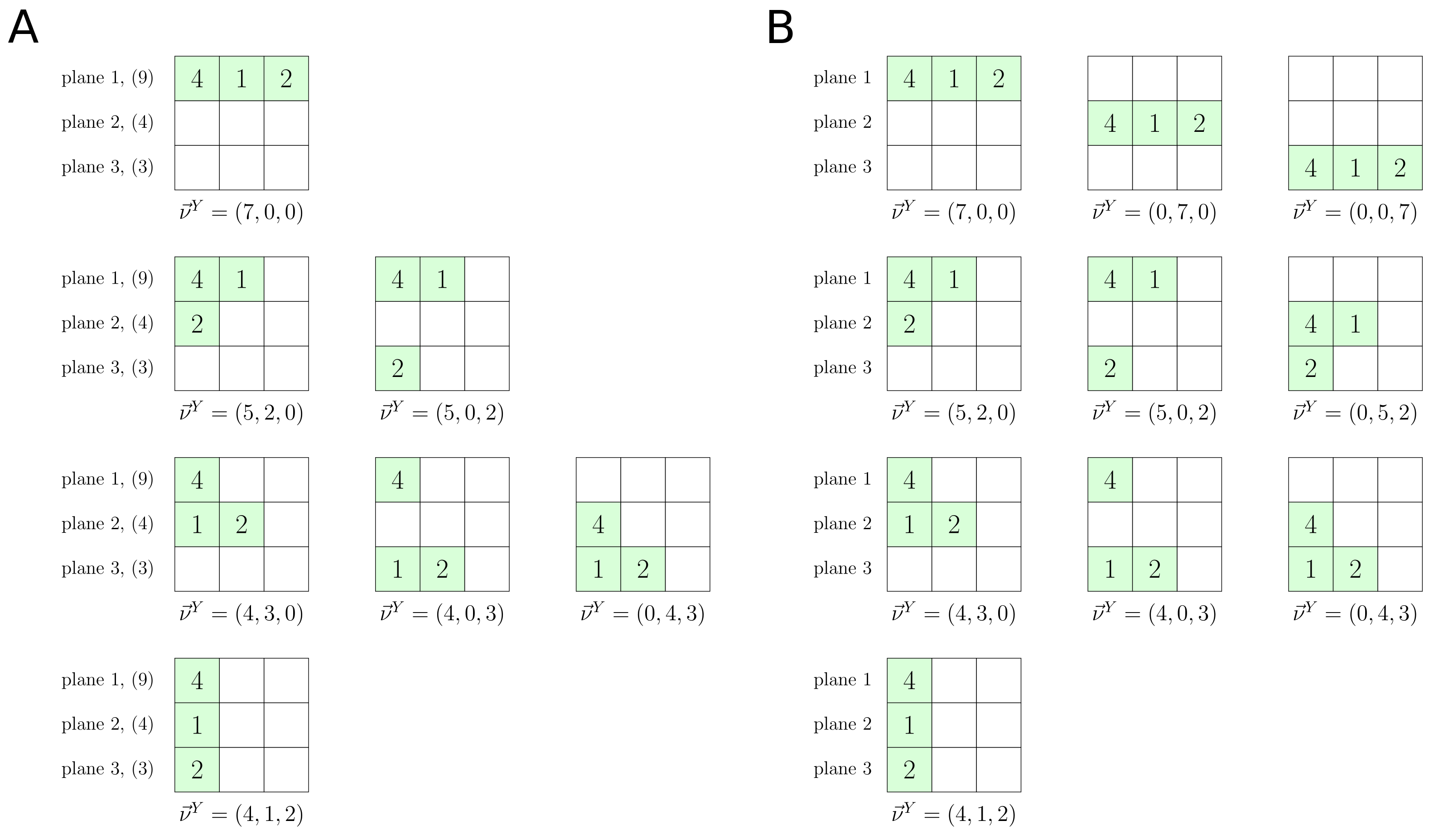}
        \caption{Illustration of summations over
        $\mathcal Y(\vec I, \mathcal N)$ and
        $\mathcal Y(\mathcal N)$ in
        Eqs.~\ref{sampling-prob-mult-planes-finite-K}
        and
        \ref{sampling-prob-mult-planes} respectively,
        for a list of allelic counts $\mathcal N = (4,1,2)$.
        (A) The finite plane case. Finite plane capacities are shown in parentheses.
        (B) The infinite plane case.}
        \label{fig:partitions}
\end{figure}

The sums in Eqs.~\ref{sampling-prob-mult-planes-finite-K} and \ref{sampling-prob-mult-planes} take into account all possible ways of sampling $n$ alleles from $M$ planes (Fig.~\ref{fig:partitions}). 
To explain these sums, let us imagine distributing $n$ books over $M$ shelves.
The books come in $k$ indivisible volume sets, and the $i$th set has
$\nu_i$ identical books in it. We would like to find all book-to-shelf arrangements,
keeping in mind that shelves have finite capacities: only $I_m$ books can be placed
on the $m$th shelf. One way to describe any book-to-shelf arrangement is to use
an $M$-dimensional vector $\vec\nu^Y$ which records how many books are placed on each shelf. For example, if $M>k$, a vector
$\vec\nu^Y = (\nu_1, ..., \nu_k, 0, ..., 0)$ with $M-k$ zeros following $k$ non-zero
entries describes placing volume sets on shelves in a particular order:
the first volume set goes on the first shelf, the second volume on the second shelf and so on
(assuming that the shelves are large enough to accommodate the volume sets),
until no more books are left, so that the remaining $M-k$ shelves remain empty.
Permutations of this arrangement, expressed as permutations of $\vec\nu^Y$ vector
elements, are also allowed (again, assuming that all the shelves are large enough).
We can also put more than one volume set on a single shelf, leading to
arrangements such as
$(\nu_1+\nu_2, \nu_3, ..., \nu_k, 0, ..., 0)$ with $M-k+1$ zero and $k-1$ non-zero entries.
As before, this arrangement is allowed only if the number of books on each shelf does not
exceed shelf capacities. Note that the question of capacity does not arise in the
infinite allele limit, since the shelves become effectively infinitely long.

In order to systematically list all the arrangements
for volume sets $(\nu_1, ..., \nu_k)$, we follow a simple rule:
if the $k$th set of $\nu_k$ books is placed on the $m$th shelf,
the $(k+1)$th set of $\nu_{k+1}$ books goes either on the same shelf
or on the $m'$th shelf with $m'>m$.
Taking elements of $(\nu_1, ..., \nu_k)$ one by one
and changing the initial shelf (onto which the 1st volume set is placed)
and the number of volume sets on each shelf, we can generate a set of all permutations
of $\vec \nu^Y$ elements. We shall call this set $\mathcal Y(\vec I, \mathcal N)$
since it depends on 
both the shelf capacities $\vec I = (I_1, ..., I_M)$ and the volume sets $\mathcal N$.
In the limit of infinite shelf capacity the dependence on shelf sizes disappears, and the
set of all permutations will be called $\mathcal Y(\mathcal N)$.
To include all possible arrangements, we need to perform the book-placing procedure
for each distinct permutation of $\mathcal N = (n_1, ..., n_k)$.

Now, if we replace shelves with fitness planes and volume sets with allelic counts,
we obtain an algorithm for generating all allowed placements of allelic counts
on fitness planes. The non-negative indices $i_1 \dots i_M$ in Eqs.~\ref{sampling-prob-mult-planes-finite-K} and \ref{sampling-prob-mult-planes} represent the number of
volume sets (allelic counts) on each shelf (fitness plane).
The distribution of alleles among fitness planes of finite capacity is illustrated in 
Fig.~\ref{fig:partitions}A for $M=3$ and a vector of allelic counts $\vec\nu = (4,1,2)$; 
the infinite-plane case is shown in Fig.~\ref{fig:partitions}B.

Next, let us consider the monomorphic limit of Eq.~\ref{sampling-prob-mult-planes}:
$\theta \to 0$ with finite $\beta$ and $\gamma$. It can be shown that
\begin{equation}
  \mathcal F(\theta \vec\gamma; \theta; \vec\beta)
  \xrightarrow[\theta\to 0]{}
  \sum_{m=1}^M \gamma_m e^{\beta_m},
\end{equation}
leading to
\begin{equation}
\begin{split}
   \mathbb P[\{n\}] &= 1+ O(\theta), \\
   \mathbb P[\{n_1, ..., n_k\}] &= O(\theta^{k-1}).
\end{split}
\end{equation}
Therefore, as expected, the $\mathbb P[\{n\}]$ ($k=1$) term predominates
in the monomorphic limit.

By construction, Eqs.~\ref{sampling-prob-2-planes} and \ref{sampling-prob-mult-planes}
reduce to the neutral limit (Eq.~\ref{sampling-prob-neutral}) when all fitness values
are the same. In addition, the neutral limit is reproduced in the strongly polymorphic,
mutation-dominated limit, defined as $\theta \to \infty$ with finite $\beta$ and $\gamma$.
In this limit,
\begin{equation}
  \mathcal F(\vec\gamma \theta + \vec \nu_Y; \theta+n; \vec\beta) \to
  \mathcal F(\vec\gamma \theta; \theta; \vec\beta),
\end{equation}
and Eq.~\ref{sampling-prob-mult-planes} reduces to the neutral result. This is expected
since selection effects become vanishingly small in this regime.


\subsection*{Frequency spectrum for arbitrary fitness landscapes}

The frequency spectrum $\Phi(x)$ is a standard way of characterizing allele frequency distributions
in evolving populations~\cite{nielsen-05}; $\Phi(x) dx$ is defined as the number of alleles in the population
with frequency in the $(x,x+dx)$ range. 
Therefore, according to Eq.~\ref{allele-frequency-distribution} the steady-state frequency spectrum is given by
\begin{equation} \label{Phi:def}
  \Phi(x) = \sum_{i=1}^K p_i(x),
\end{equation}
where
$p_i(x) = \int \prod_{j\ne i}dx_j p(\vec x)$
is the marginalized allele frequency distribution for the $i$th allele.
Note that according to Eq.~\ref{Phi:def} $\Phi(x)$ is normalized as follows:
\begin{equation}
  \int_0^1 x \Phi(x) dx = 1.
\end{equation}
Correspondingly, $x \Phi(x) dx$ is the probability that an allele randomly
drawn from the population has its frequency in the population in the $(x,x+dx)$ range. 
The frequency spectrum can be used to find $\mathbb E[k]$,
the mean number of distinct alleles in a sample of size $n$:
\begin{equation} \label{mean-number-of-alleles-via-freq-spectrum}
  \mathbb E [k] = \int_0^1 \left(1-(1-x)^n\right) \Phi(x) dx.
\end{equation}

For a landscape with $M$ distinct fitness values, the frequency spectrum
is given by (Appendix~\ref{appendix:frequency-spectrum})
\begin{equation} \label{frequency-spectrum-general}
\begin{split}
	\Phi(x) =&
	\frac{\Gamma(K\epsilon)}{\Gamma(\epsilon)\Gamma((K-1)\epsilon)}
	x^{\epsilon -1} (1-x)^{(K-1)\epsilon-1}
	\\
	&\times
	\sum_{i=1}^K
	e^{\beta_i x}
	\frac{\mathcal F (\vec\epsilon_i; (K-1)\epsilon; (1-x)\vec\beta_i)}
	{\mathcal F(\vec\epsilon; K\epsilon; \vec\beta)},
\end{split}
\end{equation}
where the $(K-1)$-dimensional vectors $\vec\beta_i$ and $\vec\epsilon_i$ are obtained from
$K$-dimensional vectors $\vec\beta$ and $\vec\epsilon$ by removing their $i$th components.
The formula above is valid for arbitrary number of alleles $K$,
mutation rate, and population size.

Using Eq.~\ref{frequency-spectrum-general}, the expected number of distinct alleles in a sample of size $n$ can be computed as
\begin{equation} \label{E-r}
\begin{split}
  \mathbb E[k]
  =&
  K-
  \frac{1}{\mathcal F(\vec\epsilon; K\epsilon; \vec\beta)}
  \sum_{i=1}^K
  \sum_{j=0}^\infty
  \frac{1}{(K\epsilon)^{(j+n)}}
  \\
  &\times
  \sum_{j_1+...+j_K=j}
  \frac{\beta_1^{j_1} ... \beta_K^{j_K}}{j_1! ... j_K!}
  \epsilon^{(j_1)} ... \epsilon^{(j_K)} ((K-1)\epsilon+j-j_i)^{(n)}.
\end{split}
\end{equation}
In the $K \to \infty$ limit (and assuming that all $M$ fitness planes
also become infinite in this limit), Eq.~\ref{frequency-spectrum-general}
simplifies to
\begin{equation} \label{freq-spectrum}
  \Phi(x) = \theta x^{-1} (1-x)^{\theta-1}
  \frac{\mathcal F (\theta \vec\gamma; \theta; (1-x)\vec\beta)}
  {\mathcal F(\theta\vec\gamma; \theta; \vec\beta)}
  \sum_{m=1}^M \gamma_m e^{\beta_m x}.
\end{equation}
Furthermore, in the case of two fitness states ($M=2$) we can simplify
Eq.~\ref{freq-spectrum} using Eq.~\ref{mathcalF=F11}~\cite{ewens-80}:
\begin{equation} \label{Phi:twoplanes}
  \Phi(x) = \theta x^{-1} (1-x)^{\theta-1}
    ( 1-\gamma + \gamma e^{\beta x} )
    \frac{\f11(\gamma \theta; \theta; (1-x)\beta)}
    {\f11(\gamma \theta; \theta; \beta)}.
\end{equation}

For neutral evolution, we set all $\beta_i$ to $0$;
Eq.~\ref{freq-spectrum} then yields
\begin{equation}
  \Phi_{\mathrm{neutral}}(x) = \theta x^{-1} (1-x)^{\theta-1},
\end{equation}
which looks like two-allele steady-state allele frequency distribution.
In the strongly monomorphic limit and the absence of selection, the steady-state distribution (Eq.~\ref{allele-frequency-distribution}) simplifies to (Eq.~\ref{strongly-monomorphic-limit}):
\begin{equation}
	p_{\mathrm{neutral}}(x) = \frac 12[\delta(x)+\delta(x-1)].
\end{equation}
Since $\Phi(x) = K p(x)$, we obtain
\begin{equation}
	\Phi_{\mathrm{neutral}}(x) = \frac K2 [\delta(x)+\delta(x-1)]
\end{equation}
in the monomorphic limit of neutral evolution.

Using Eq.~\ref{mean-number-of-alleles-via-freq-spectrum},
we can obtain a standard expression for the mean number of alleles observed in neutral
evolution~\cite{ewens-04,li-78,ewens-80}: 
\begin{equation} \label{Ek:neutral}
  \mathbb E[k] = \theta \left(\psi(\theta+n)-\psi(\theta)\right)
  = \theta \sum_{i=0}^{n-1} \frac{1}{i+\theta},
\end{equation}
where $\psi(z) = \Gamma'(z)/\Gamma(z)$ is the digamma function. Note that
$\mathbb E[k] \to 1$ in the monomorphic limit ($\theta \to 0$) and 
$\mathbb E[k] \to n$ in the strongly polymorphic limit ($\theta \to \infty$).

Finally, we observe that Eq.~\ref{Ek:neutral} can also be derived by setting all $\beta_i = 0$ in
Eq.~\ref{E-r}:
\begin{equation}
\begin{split}
  \mathbb E[k]
  &=
  K-
  \sum_{i=1}^K
  \frac{\Gamma(K\epsilon)}{\Gamma((K-1)\epsilon))}
  \frac{\Gamma((K-1)\epsilon+n))}{\Gamma(K\epsilon +n)}
  \\
  &\xrightarrow[K \to\infty]{}
  K - K \frac{1-\psi(K\epsilon)\epsilon+ o(\epsilon^2)}
  {1-\psi(K\epsilon+n)\epsilon +o(\epsilon^2)}
  \\
  &=
  \theta (\psi(\theta+n)-\psi(\theta)).
\end{split}
\end{equation}

In the monomorphic limit ($\theta \to 0$), Eq.~\ref{freq-spectrum} becomes
\begin{equation}
  \Phi_\mathrm{mono}(x) = \theta x^{-1} (1-x)^{-1}
  \frac{\sum_{m=1}^M \gamma_m e^{x\beta_m} \sum_{m=1}^M \gamma_m e^{(1-x)\beta_m}}
  {\sum_{m=1}^M \gamma_m e^{\beta_m}}.
\end{equation}
In this limit, $x \Phi_\mathrm{mono}(x)$ is non-zero only
if $x \simeq 1$, where $\Phi_\mathrm{mono}(x) \simeq \theta x^{-1} (1-x)^{-1}$.
Note that selection effects disappear: the entire population is in the same allelic
state due to genetic drift, performing a random walk on the uppermost fitness plane.

In the strongly polymorphic, mutation-dominated limit ($\theta \to \infty$), Eq.~\ref{freq-spectrum}
simplifies to
\begin{equation}
  \Phi_\mathrm{poly}(x) = \theta x^{-1} (1-x)^{\theta}
  e^{-x\sum_{m=1}^M \gamma_m \beta_m}
  \sum_{m=1}^M \gamma_m e^{x\beta_m}.
\end{equation}
In this case, $x \Phi_\mathrm{poly}(x)$ is peaked around 
$x \simeq 0$, with $\Phi_\mathrm{poly}(x) \simeq \theta x^{-1} (1-x)^{\theta}$.
The population is completely delocalized, with each member of the population
in a distinct allelic state and negligible selection effects. This is not surprising
since mutation dominates both selection and genetic drift in this limit.

\subsection*{Fitness landscape models}

We have carried out validation
of our theoretical predictions against numerical simulations.
We have used the Moran model of population genetics~\cite{ewens-04,moran-58} to evolve 
a population of $N = 10^3$ haploid organisms, each of which could be in one of $K$ allelic states.
Specifically, at each step a parent is chosen by randomly sampling the population
with weights proportional to the fitness of each individual. An offspring
is then produced as an exact copy of the parent. Next, the offspring
undergoes mutation with the probability $\mu$. Finally, the population is uniformly
sampled to choose an organism that will be replaced by the offspring, keeping the
overall population size constant.
Probabilities of sampling $n$ individuals from the population were calculated as
averages over $10^6$ samples gathered from $10^3$ independent
runs.  For each run, a randomly generated initial population was evolved
to steady state, after which $n$ individuals were sampled from the
population with replacement $10^3$ times, waiting $\sim 1/\mu$ generations
between samples.

We consider two types of models with different mutational moves. In the first model, each allele is allowed to mutate into any of the other $K-1$ alleles with equal probabilities. We call this model fully-connected (FC); it corresponds to our theory which was developed for FC networks.
In the second model, a more realistic move set of single-point mutations
is implemented: each organism's genome is represented by a sequence of integers $a_1 ... a_L$
of length $L$, where $0 \le a_i \le A-1$. A mutation replaces an integer
at a randomly chosen site with one of the remaining $A-1$ integers; all the replacements have equal probabilities. We call this model a
single-point mutation (SPM) model.

Finally, we assign a fitness value to each allele. We focus on the landscapes
in which alleles can have either low or high fitness values (the ``two-plane'' model),
or low, intermediate, and high fitness values (the ``three-plane'' model).
The fractions of alleles found in each plane are given by $\vec\gamma$:
$\vec\gamma = (\gamma, 1-\gamma)$ for the two-plane model and
$\vec\gamma = (\gamma_1, \gamma_2, 1-\gamma_1-\gamma_2)$ for the three-plane model.
In the FC model, the mutational neighborhood of each allele is the same,
so that any desired allele fractions $\vec\gamma$ can be implemented. However, in the 
SPM model the fractions of neutral, beneficial and deleterious moves in each
plane will depend on $\vec\gamma$ and the assignment of states to planes. We wished to produce
non-trivial distributions of neutral moves on the fitness planes, with
mutational neighborhoods of some alleles being completely neutral in each
plane.  
Another condition was that the number of alleles in each plane should
decrease with its fitness, to reflect the fact that higher-fitness solutions
are harder to find.

To fulfill these requirements, we chose to assign fitness values in the SPM model in the following way. We use the sequence length $L=10$ and the alphabet size $A=4$.
For each sequence $a_1 ... a_L$ we compute a score $z = a_1 + ... + a_L$.
We compare these scores with a set of cutoffs $(c_1, ..., c_{M-1})$ for the $M$-plane
landscape. For the two-plane landscape, the fitness is $1$
if $z \le c_1$, and $1+s$ otherwise. We use the cutoff $c_1 = 17$,
which yields $\vec\gamma = (0.758,0.242)$. For the three-plane landscape,
if $z \le c_1$ the fitness is $1$, if $c_1 < z \le c_2$ the fitness is $1+s-\Delta s$, and if $z > c_2$ the fitness is $1+s+\Delta s$.
We choose the cutoffs $c_1 = 17$ and $c_2 = 21$,
which lead to $\vec\gamma = (0.758, 0.210, 0.032)$. In order to compare
FC and SPM simulations directly, we use the same values of $\vec\gamma$ in the corresponding FC models.

Note that in the neutral case the exact mapping between $\theta$ and $\mu$ is given
by $\theta = {N \mu}/{(1 - \mu)}$ for the Moran model.~\cite{ewens-04}
However, it is unclear if this mapping can be extended to the non-neutral cases
considered here. In any event, for the population size and the values of $\theta$
investigated below, $\mu = {\theta}/{(N + \theta)} \simeq {\theta}/{N}$.
Therefore, we use the diffusion theory result $\theta = N\mu$ in comparing
theoretical predictions with numerical simulations.

\subsection*{The effective population size approximation}

In the monomorphic limit, we expect the effective population size (EPS) approximation
to hold~\cite{charlesworth-93,desai-12}: population dynamics is neutral but with the rescaled
population size $N^*$. Indeed, in the two-plane case Eq.~\ref{sampling-prob-2-planes} reduces to

\begin{equation} \label{sampling-prob-neutral:Neff}
\mathbb P[\{n_1, ..., n_k\}] \xrightarrow[\theta\to0]{}
\frac{N_P}{k!}
 \frac{n!}{\prod_{i=1}^k n_i}
 \theta^{k-1} (1-\gamma)^{k-1}
\end{equation}
in the $\theta \to 0$ limit, which corresponds to
the $s \gg \mu$ regime when $\beta$ is finite; Eq.~\ref{sampling-prob-neutral:Neff} is the same as the neutral sampling formula (Eq.~\ref{sampling-prob-neutral}) in the monomorphic
limit if the population size is rescaled: $N \to N^* = (1-\gamma) N$.
This result can be generalized to the landscape with multiple fitness planes, in which
case
\begin{equation}
	N^* = \gamma_m N,
\end{equation}
where $\gamma_m$ is a fraction of nodes with the highest fitness.

However, the EPS approximation breaks down in the polymorphic regime.
Indeed, if we take the $N \to \infty$ limit, which keeps $\beta/\theta$ finite
(i.e., the ratio of selection and mutation forces remains finite as population size increases),
it can be shown for the two-plane landscape that
\begin{equation} \label{strongly:polymorphic:sample}
	\frac{\mathbb{P}[\{n_1, ..., n_k\}]}
	{\mathbb{P}[\{n_1, ..., n_k\}|\beta=0]}
	=
	\sum_{m=0}^\infty c_m \left( \frac{s}{\mu} \right)^m \equiv S
\end{equation}
where $\mathbb{P}[\{n_1, ..., n_k\}|\beta=0]$ is given by Eq.~\ref{sampling-prob-neutral}, and coefficients $c_m$ depend solely on $n_1, ..., n_k$. Since the right-hand side of 
Eq.~\ref{strongly:polymorphic:sample} does not depend on the population size,
it can be used to define $N^* = S^{1/(k-n)} N$. However, this definition will be
sample-specific, due to dependence of $S^{1/(k-n)}$ on $n_1, ..., n_k$. Thus there is
no global rescaling of the population size in the strongly polymorphic regime,
and evolutionary dynamics is non-neutral~\cite{desai-12}.





\begin{figure}[!t]
        \centering
        	\includegraphics[width=\textwidth]{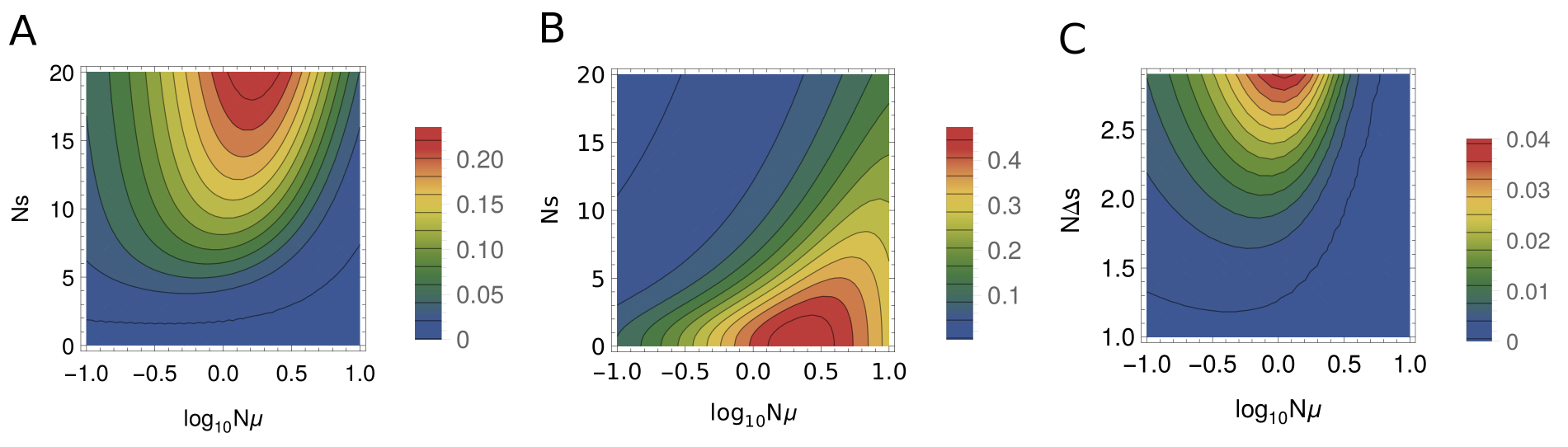}
        \caption{Probabilities of all possible partitions of $n = 3$ alleles ($\{3\}$, $\{2,1\}$ 	
        $\{1,1,1\}$) sampled from the population
        of size $N = 10^3$. (A) and (B) KL divergences for the two-plane fitness landscape
        as a function of the mutation rate $N\mu$ and the selection
        coefficient $Ns$ scaled by the population size, for partition probabilities with and without selection (A), and partition probabilities with selection
        compared with the EPS approximation (Eq.~\ref{sampling-prob-neutral:Neff}) (B).
        (C) KL divergences for the sampling probabilities of all possible partitions
        on a three-plane vs. two-plane landscape. 	
        Alleles in the three planes have 
        fitnesses $1$, $1 + s - \Delta s$ and
        $1 + s - \Delta s$ respectively, with $Ns = 6$ for both two and three-plane landscapes.
        }
        \label{fig:dev-2-3-planes}
\end{figure}

\subsection*{Detection of selection signatures}

\begin{figure}[tb]
	\centering
	\includegraphics[width=\textwidth]{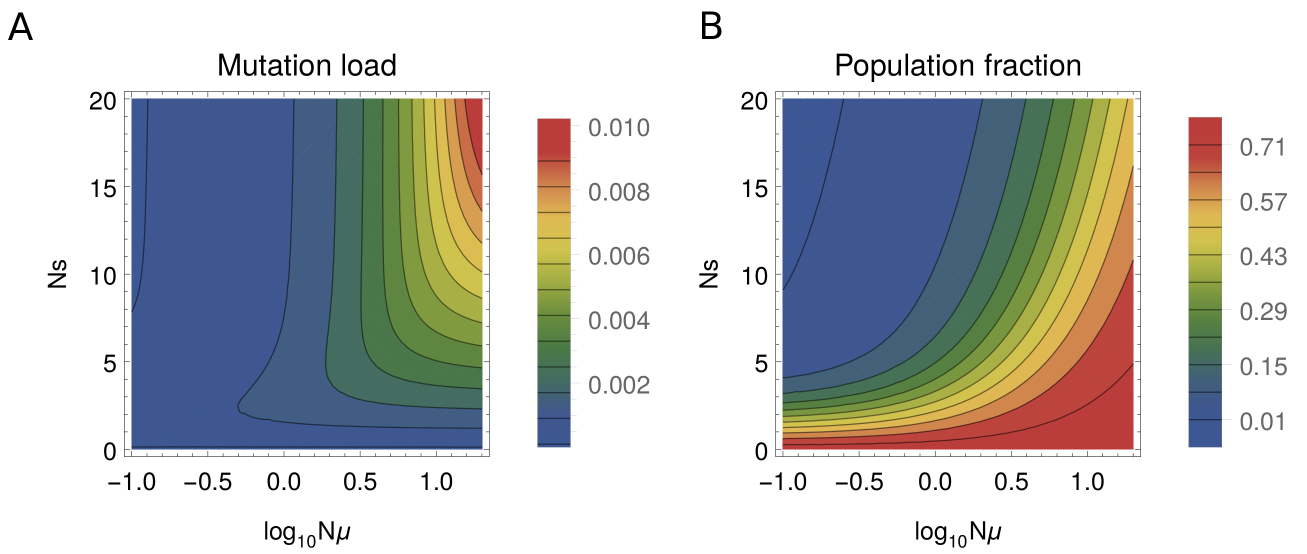}
	\caption{(A) Mutation load (Eq.~\ref{L:two-plane}) and (B) the population fraction
	on the lower plane (Eq.~\ref{fraction.lower.plane:two-plane}) for the two-plane fitness landscape, as a function of the mutation
	rate ($N\mu$) and the selection strength ($Ns$) rescaled by the population size.}
	\label{fig:mutload}
\end{figure}

As discussed above, in general we expect allele diversity to deviate from neutrality, making it possible to detect selection signatures using sequences sampled from
a population as input. To investigate non-neutral population
dynamics, we compute probabilities for all partitions $\{n_1, ..., n_k\}$ of $n$
alleles sampled from the population evolving under selection, and compare 
them with steady-state partition probabilities obtained
under neutral evolution and the EPS approximation.

We use the Kullback-Leibler (KL) divergence to quantify the difference between two probability distributions~\cite{kullback-51}:
\begin{equation}
	KL(P||Q) = \sum_i P(i) \log \frac{P(i)}{Q(i)}.
\end{equation}

For the two-plane system, we first compare
partition probabilities under selection, $P (i) = \mathbb{P}[\{n_1, ..., n_k\}|\theta, \beta]$, with the corresponding neutral probabilities,
$Q (i) = \mathbb{P}[\{n_1, ..., n_k\}|\theta, \beta=0]$. Here, $i$ labels distinct
partitions.
In Fig.~\ref{fig:dev-2-3-planes}A, we plot the KL divergence as a function of the
mutation rate and the selection strength for the two-plane fitness landscape. We observe that
evolutionary dynamics is essentially neutral if selection is weak ($s \le \mu$);
in addition, the range of selection coefficients for which neutrality holds increases in
the monomorphic regime ($N \mu \leq 1$). On the other hand,
population statistics is clearly non-neutral when the population is polymorphic
and the separation between the two fitness planes is large.
Next, we compute the KL divergence $KL(P||Q^*)$ between the EPS probability
distribution, $Q^* (i) = \mathbb{P}[\{n_1, ..., n_k\}|\theta^*, \beta=0]$, where $\theta^* = (1 - \gamma) \theta$, and $P (i)$ (Fig.~\ref{fig:dev-2-3-planes}B). We see that the EPS approximation fails
in the polymorphic, weak-selection regime. Overall, the neutral and EPS approximations
are approximately complementary -- for example, in the strong-selection ($s \gg \mu$), polymorphic regime, when evolutionary dynamics becomes non-neutral, it is well approximated by the EPS model.


In Fig.~\ref{fig:dev-2-3-planes}C we show KL divergences between partition probability distributions on two- and three-plane fitness landscapes. We observe that
the partition probabilities are essentially two-plane (i.e., there are no
selection signatures indicating presence of intermediate-fitness alleles) if the
population is monomorphic ($N \mu \leq 1$), or if the distance between the two
upper planes is smaller than the mutation rate ($\Delta s \le \mu$). However, there is
a considerable parameter region in which deviations between two and three-plane
sampling probabilities appear to be significant (with KL divergences between the two distributions of 0.01 or more), making it possible
to detect three distinct fitness states in the sampling data.



\subsection*{Mutation load}

By definition, the mutation load is given by~\cite{Gillespie2004,Crow1970}
\begin{equation}
	L = \frac{f_\mathrm{max}-\langle f \rangle}{f_\mathrm{max}},
\end{equation}
where $f_\mathrm{max}$ is the maximum fitness, and
$\langle f \rangle = \sum_{i=1}^K x_i f_i$ is the mean population fitness.
To estimate the mutation load at steady state, we compute the expected value
of the mean population fitness over multiple realizations of the stochastic process:
\begin{equation}
	\mathbb{E}[ \langle f \rangle] = \sum_{i=1}^K \mathbb{E}[x_i] f_i
	=
	f_K+\frac{1}{N}\sum_{i=1}^{K-1} \mathbb{E}[x_i]\beta_i
	=
	f_K+\frac{1}{N}\sum_{i=1}^{K-1} \frac{\partial \log Z}{\partial \log \beta_i},
\end{equation}
where $Z = B(\vec\epsilon) \mathcal{F}(\vec\epsilon; |\vec\epsilon|; \vec\beta)$
is the normalization in the
steady-state allele frequency distribution (Eq.~\ref{allele-frequency-distribution}).
Choosing $f_K$ to be the maximum fitness, we obtain the following expression for the mutation load:
\begin{equation}
	L = -\frac{1}{N f_K} \sum_{i=1}^{K-1} \mathbb{E}[x_i]\beta_i
	=
	-\frac{1}{N f_K} \sum_{i=1}^{K-1} \frac{\partial \log Z}{\partial \log \beta_i}.
	\label{L:general}
\end{equation}

For the two-plane system, Eq.~\ref{L:general} yields (Appendix~\ref{appendix:sampling-two-planes}):
\begin{equation}
	L = - \frac{\beta\gamma}{N (1+s)}
		\frac{\f11(\gamma\theta+1; \theta+1; \beta)}{\f11(\gamma\theta; \theta; \beta)}.
	\label{L:two-plane}
\end{equation}
Note that in the two-plane system
\begin{equation}
	\mathbb{E}[x_{\mathrm{low}}] = \frac{1 + s}{s} L,
	\label{fraction.lower.plane:two-plane}
\end{equation}
where $\mathbb{E}[x_{\mathrm{low}}]$ is the average fraction of the population on the
lower plane.

\begin{figure}[!tb]
        \centering
        	\includegraphics[width=\textwidth]{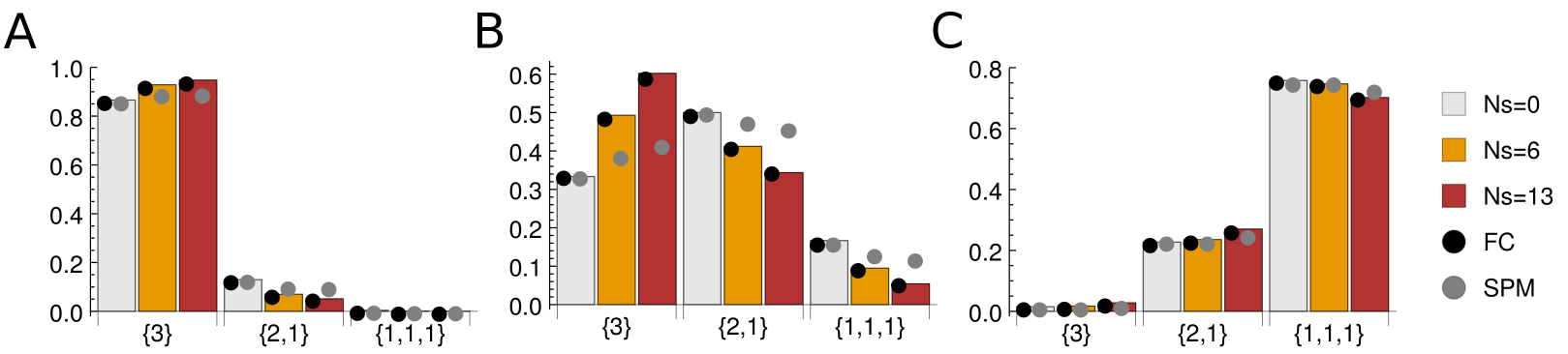}
        \caption{Partition probabilities for the two-plane fitness landscape.
        Shown are sampling probabilities of all partitions with $n=3$:
        $\{3\}$, $\{2,1\}$, $\{1,1,1\}$.
        	Bars: theoretical predictions in the infinite allele limit.
        	Black circles: numerical simulations on the FC sequence network.
        	Grey circles: numerical simulations on the SPM sequence network.
        	In all simulations, alphabet size $A = 4$, sequence length $L =10$,
       	and population size $N = 10^3$ were used.
       	Partition probabilities were estimated from 
       	$10^6$ samples as described in the main text.
        	(A) Monomorphic population, $N\mu = 0.1$.
        	(B) Weakly polymorphic population, $N\mu = 1.0$.
        	(C) Strongly polymorphic population, $N\mu = 10.0$.
        	Note that the corresponding KL divergences are listed in Table~\ref{table:KL}.
        }
        \label{fig:two-infinite-planes}
\end{figure}

Mutation loads for the two-plane fitness landscape are shown in Fig.~\ref{fig:mutload}A
over a range of selection strengths and mutation rates. As expected,
we observe that the largest deviations from the maximum fitness occur in the
strong-mutation, strong-selection regime, where a fraction of the population
is constantly displaced to the lower plane by mutation, incurring a fitness
cost. Correspondingly, at a given value of selection strength the mutation load
increases with the mutation rate. In the monomorphic regime the mutation load
is vanishingly low because the entire population condenses to a single allelic state and moves randomly on the upper plane.
The fraction of the population on the lower fitness plane is shown in Fig.~\ref{fig:mutload}B.
The fraction is high when the separation between the two planes is low and, at a fixed
separation, it increases with the mutation rate.

\begin{figure}[!tb]
	\centering
	\includegraphics[width=\textwidth]{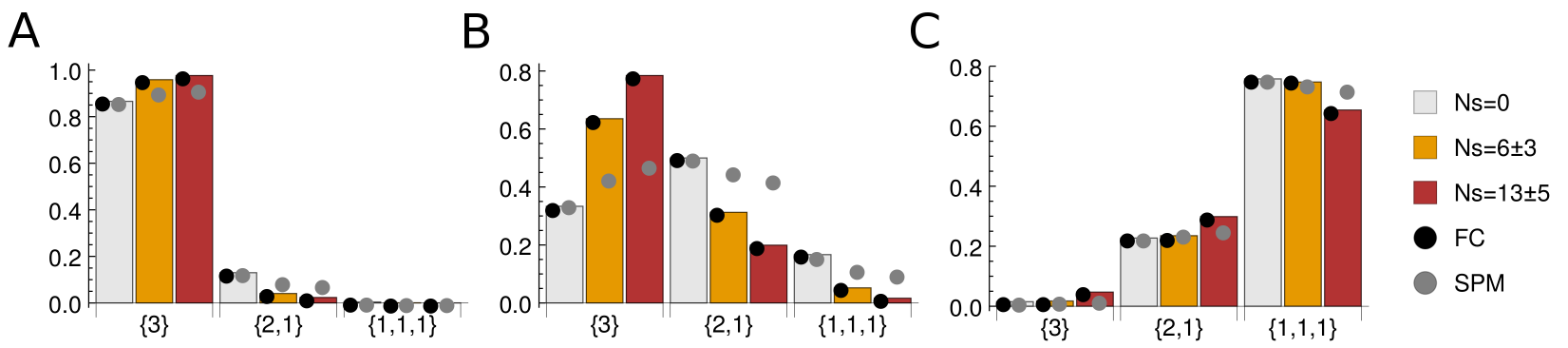}
        \caption{Partition probabilities for the three-plane fitness landscape.
        	All parameters and symbols are as in Fig.~\ref{fig:two-infinite-planes}, unless indicated otherwise; KL divergences are listed in Table~\ref{table:KL}.
        }
	\label{fig:multiple-planes}
\end{figure}



\subsection*{Partition probabilities on fully-connected vs. single-point-mutant networks}

Our theoretical results have been developed for fully-connected networks in which an allele can mutate into any other allele.
However, this model is not realistic for protein or nucleotide sequences, in which
mutational neighborhoods of a given sequence consist of single-point mutants, i.e. sequences that differ from each other at
only a single site. Here we investigate how partition probabilities change if we
switch from the FC to the SPM allele network described above. In Fig.~\ref{fig:two-infinite-planes},
we compare theoretical predictions with numerical simulations on the
FC and SPM networks in the two-plane system. Overall, we observe excellent agreement between
theory and simulations on FC networks.
Furthermore, we see that the agreement between 
SPM simulations and our theoretical results is reasonable: in nearly all cases, the
predicted ranking of the sample partitions, as well as the ranking within any given
sample partition with respect to $Ns$,
are preserved. The largest discrepancies occur
in the weakly polymorphic ($N \mu = 1$), strong-selection regime ($Ns = 6,13$).

The situation is qualitatively similar when a three-plane fitness landscape
is considered (Fig.~\ref{fig:multiple-planes}). We again observe excellent agreement between theory and FC simulations
and, overall, reasonable agreement between theory and SPM simulations, with the
largest discrepancies again occurring in the weakly polymorphic, strong-selection  regime.

\begin{table}[hbt!]
\begin{center}
\caption{KL divergences between theoretical predictions and numerical simulations for the two-plane fitness landscape (Fig.~\ref{fig:two-infinite-planes}) and the three-plane fitness landscape (Fig.~\ref{fig:multiple-planes})).}
\label{table:KL}
\begin{tabular}{|l|l|c|c|c|c|c|c|}
\hline
\multicolumn{2}{|c|}{} & \multicolumn{3}{c|}{\textbf{Two-plane landscape}}                             & \multicolumn{3}{c|}{\textbf{Three-plane landscape}}                             \\ \cline{3-8} 
\multicolumn{2}{|c|}{}                  & $Ns=0$            & $Ns=6$            & $Ns=13$           & $Ns=0$            & $Ns=6\pm3$        & $Ns=13\pm5$       \\ \hline \hline
{$N\mu=0.1$}    & \textbf{FC}     & $7\times 10^{-7}$ & $3\times 10^{-5}$ & $8\times 10^{-5}$ & $1\times 10^{-5}$ & $2\times 10^{-6}$ & $7\times 10^{-8}$ \\ \cline{2-8} 
                               & \textbf{SPM}    & $4\times 10^{-5}$ & $8\times 10^{-3}$ & $2\times 10^{-2}$ & $2\times 10^{-6}$ & $2\times 10^{-2}$ & $3\times 10^{-2}$ \\ \hline
{$N\mu=1$}      & \textbf{FC}     & $5\times 10^{-5}$ & $2\times 10^{-5}$ & $1\times 10^{-4}$ & $3\times 10^{-5}$ & $4\times 10^{-5}$ & $5\times 10^{-6}$ \\ \cline{2-8} 
                               & \textbf{SPM}    & $4\times 10^{-5}$ & $2\times 10^{-2}$ & $8\times 10^{-2}$ & $2\times 10^{-4}$ & $9\times 10^{-2}$ & $2\times 10^{-1}$ \\ \hline
{$N\mu=10$}     & \textbf{FC}     & $9\times 10^{-6}$ & $8\times 10^{-6}$ & $2\times 10^{-5}$ & $7\times 10^{-6}$ & $1\times 10^{-4}$ & $2\times 10^{-5}$ \\ \cline{2-8} 
                               & \textbf{SPM}    & $4\times 10^{-5}$ & $2\times 10^{-4}$ & $3\times 10^{-3}$ & $6\times 10^{-6}$ & $1\times 10^{-4}$ & $2\times 10^{-2}$ \\ \hline
\end{tabular}
\end{center}
\vspace{-0.6cm}
\end{table}

\subsection*{Network size effects}

\begin{figure}[tb]
        \centering
        \begin{subfigure}[b]{0.75\textwidth}
                \includegraphics[width=\textwidth]{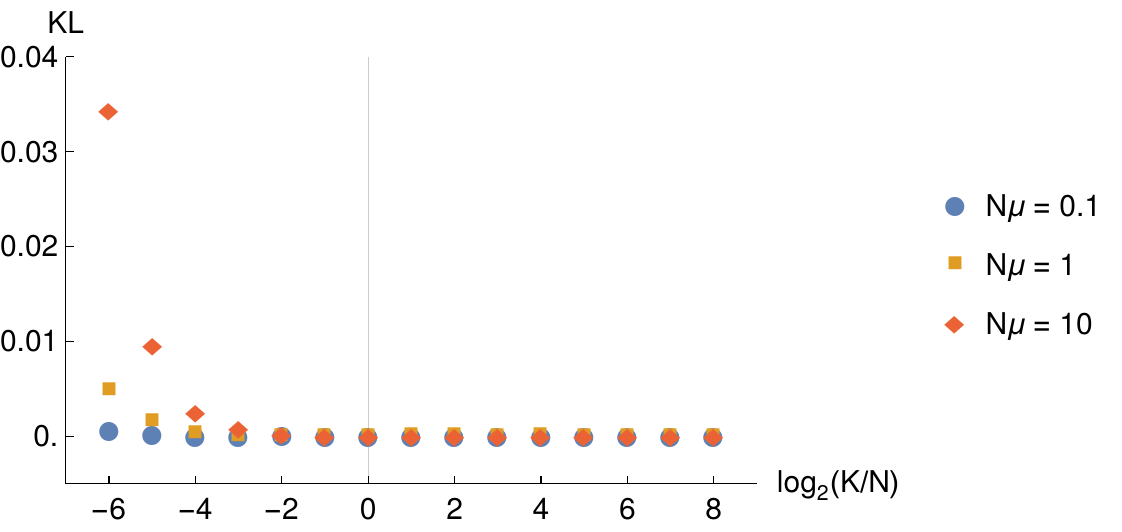}
        \end{subfigure}
        \caption{KL divergence between computational and theoretical partition probabilities on the FC two-plane fitness landscape ($Ns = 6$, $\vec\gamma = (0.758,0.242)$), as a function of the log ratio between the total number of alleles $K$ and the population size $N$.
The sample size is $n=3$; partition probabilities were estimated from 
$10^6$ samples as described in the main text.
Population size is $N = 10^3$, and the total number of alleles is $K = 10^3 \times 2^i$, $i \in \{-6 \dots 8\}$. For smaller networks, the number of the nodes in the upper and lower planes
had to be rounded to the nearest integer.
Diamonds: polymorphic population ($N\mu = 10.0$), squares: weakly polymorphic population ($N\mu = 1.0$), circles: monomorphic population ($N\mu = 0.1$).
        The solid vertical line indicates the case of the network size equal to the population size ($K=N$). 
        }
        \label{fig:var-K}
\end{figure}

Although our approach is valid for an arbitrary number of alleles $K$,
statistics of allele diversity in a population under selection become substantially easier to deal with 
in the infinite-allele limit. As discussed in the Introduction, this limit is
justified since our focus here is on evolution of protein, RNA and DNA sequences,
where the number of alleles grows exponentially with sequence length. Nonetheless, we have systematically investigated the extent of deviations between our infinite-allele theory results and simulations
as the number of alleles $K$ decreases and becomes comparable to the population size $N$. Fig.~\ref{fig:var-K}
shows the KL divergence between partition probabilities derived theoretically for the
two-plane landscape in the infinite-allele
limit (Eq.~\ref{sampling-prob-2-planes}) and obtained numerically on finite-size FC networks. We consider three regimes:
monomorphic ($N \mu = 0.1$), weakly polymorphic ($N \mu = 1.0$),
and strongly polymorphic ($N \mu = 10.0$). In the latter two cases,
noticeable deviations between theory and simulations begin to appear below the 
$K \sim N$ regime; the agreement improves as the population becomes more monomorphic.
We conclude that our theory is applicable over a wide range of mutation
rates, as long as the network size is comparable to, or greater than,
the population size.

\section*{Discussion}


One of the most challenging problems in evolutionary biology is to understand evolutionary dynamics of molecular loci, such as protein or RNA-coding sequences,
or gene regulatory regions. The number of nucleotides at these loci, $L$, is  
large enough so that the total number of possible sequences, $K = A^L$, is astronomical,
far exceeding the population size $N$. Under these
conditions the evolution of a molecular locus, assumed to be decoupled by
recombination from the rest of the genome, reaches a ``de-labelled'' steady state characterized
by mutation-selection-drift balance. The allelic diversity in the population
is determined by the balance of forces of selection and drift on one hand, and
mutation on the other. The former act to reduce allelic diversity, while
the latter acts to increase it. As a result, 
population statistics such as the mean number of distinct alleles,
or the probability of seeing a certain allelic configuration in a sample, do not change with time, even though new genotypes continue to be explored on the effectively infinite
allelic network.

The neutral allelic diversity in such a system (that is, when all alleles have
the same fitness) was explored by Ewens~\cite{ewens-04,ewens-72}. The main result
of that study, the Ewens sampling formula, is widely used in population genetics.
The neutral landscape is a single plane,
with each allele connected to the other $K-1$ alleles. However, recent high-throughput
studies connecting protein sequences with phenotypes reveal
a more complex picture: generally, a functional protein such as an enzyme can be 
disrupted by a subset of mutations at each of its sites (e.g., through substitution
of a hydrophobic residue for a hydrophilic one in the protein core). Other mutations
do not significantly change protein stability, binding affinity or specificity, and are
therefore effectively neutral. Occasionally, a mutation is found which increases
the protein's fitness, but these mutations are generally infrequent. Overall,
recent experimental studies point to fitness landscapes comprised of
multiple interconnected planes. The simplest landscape of this kind has just two distinct fitness states, with functional sequences on the upper plane and non-functional sequences
on the lower plane~\cite{Podgornaia2015}. Multiple-plane fitness landscapes are characterized by extensive epistasis, which is likely to be pervasive in molecular
evolution~\cite{Lunzer2005,Romero2009,Lunzer2010,Breen2012}.

Since molecular evolution may be described by steady-state dynamics on multiple-plane fitness landscapes, it is of great interest to generalize the Ewens
sampling formula to arbitrary fitness landscapes, and to the multiple-plane class of
landscapes in particular. Tractable expressions for sampling probabilities would
enable inference of selection coefficients, relative plane sizes, and mutation rates, using DNA, RNA or protein sequences sampled from the population as input.
Here we report an extension of the Ewens sampling formula to arbitrary fitness landscapes, focusing especially on the multiple-plane case which yields
substantial simplifications in the infinite allele limit.
Unlike current state-of-the-art techniques based on the Poisson random field framework~\cite{sawyer-92}, such as the sampling probability formulas developed by
Desai et al.~\cite{desai-12}, our approach is capable of treating epistasis. However,
the essential drawback of the Ewens sampling formula and its generalizations is the ``fully-connected'' assumption (i.e., that each allele can mutate into every other allele). Moreover, the sampling formula becomes intractable for large sample sizes, due to 
a large number of terms to sum over.

Therefore, in order to study the limits of applicability of our theory, we have
carried out extensive comparisons with numerical simulations on multiple-plane
fitness landscapes. First, we checked the full-connectivity assumption inherent in the
Ewens approach by comparing
the sampling probabilities of our theory with those obtained by simulation of steady-state 
populations evolving on single-point-mutant networks. We find that the agreement,
although dependent on the details of the fitness landscape model, the values of selection coefficients, and mutation rates
(and least reliable in the weakly polymorphic regime), remains strong enough overall
to encourage application of our theoretical results to sequence data.
Note that our model of the fitness landscape was constructed specifically to create
a non-trivial distribution of neutral, deleterious and beneficial single-point
mutations for the alleles, making it in some sense as distant from the fully connected network
as possible. Thus we expect the deviations to be smaller (or at least not much worse) in natural systems.
Second, we have checked the infinite-allele assumption by systematically
reducing the number of alleles until it became lower than the population size.
We find that, over a wide range
of mutation rates, deviations between theory and simulations become significant only when the number of alleles approaches the population size from above. Thus our
assumption of the infinite network size is justified for all loci that are long enough,
such as those encoding transcribed or regulatory regions.

Robust inference of selection coefficients and mutation rates on the basis of a sample 
of population allelic states requires statistics of allelic diversity to deviate substantially from both the neutral expectation and the effective population size
(EPS) approximation. Clearly, no inference of selection signatures is possible on the
basis of limited sample data if population dynamics is close to neutral. On the other hand, in the EPS
limit only the relative size of the highest-fitness plane can be inferred. By scanning over a wide range of selection coefficients and mutation rates on
a two-plane fitness landscape, we have found that, although regions of neutral and EPS dynamics are roughly complementary, there are areas of parameter space characterized by deviations from both. Thus the use of our generalized Ewens sampling formula, which is
valid throughout the entire parameter space, is necessary for inferring selection signatures from data. Moreover, allelic diversity generated by steady-state evolutionary dynamics on a three-plane fitness landscape is sufficiently distinct from its two-plane counterpart in
the strong-selection, weakly polymorphic regime, opening up a possibility of
inferring multiple selection coefficients from the data.
Another hallmark of non-neutral population dynamics is de-localization of the
population to multiple fitness planes. With a two-plane landscape,
we expect the fraction of the population on the lower plane to increase with the
mutation rate and decrease with the distance between the two planes.
Our investigation of the mutation load confirms these predictions.

In summary, we have generalized the Ewens sampling formula to evolutionary
dynamics under selection. Although in principle our results are valid for arbitrary
fitness landscapes, focusing on the infinite allele limit and landscapes with 
just two or three distinct fitness states yields substantial simplifications,
making our approach computationally tractable and thus applicable to inferring selection signatures from high-throughput sequence data. Such multiple-plane fitness landscapes are consistent with recent large-scale studies of molecular
phenotypes~\cite{Podgornaia2015,Romero2009,Lind2010,Hietpas2011}. Unlike previous approaches, we do not assume the absence of epistasis, which
is likely to be prevalent in molecular evolution~\cite{Lunzer2005,Romero2009,Lunzer2010,Breen2012}. However, we do make
the infinite allele assumption, and, as in the Ewens original formula~\cite{ewens-72}, 
assume that each allele can mutate into any other allele. We check our theory
against numerical simulations in model systems where these
assumptions are relaxed, and find that our predictions remain accurate
enough to enable inference of evolutionary parameters from sequencing data.

\section*{Acknowledgements}

PK acknowledges financial support from a research fellowship awarded by
the Department of Physics and Astronomy, Rutgers University.
AVM was supported in part through a collaboration with Los Alamos
National Lab (LANL-DOE 20150236ER).

\section*{Data Availability}

Software and models used in this study are freely available upon
request.




\appendix

\section*{Appendix}

\section{Simplification of the sampling formula in the two-plane system} \label{appendix:sampling-two-planes}

In the two-plane system, the fitness vector has the following structure:
\begin{equation}
	\vec\beta = (\underbrace{\beta, ..., \beta}_{I}, \underbrace{0, ..., 0}_{K-I})
\end{equation}
with $I$ nonzero entries followed by $K-I$ zeros. In this case, Eq.~\ref{gen-hyp-def}
involves summation over only $I$ indices:

\begin{equation}\label{mathcalF=F11}
\begin{split}
  \mathcal F(\vec\epsilon; \theta; \vec\beta)
  &=
  \sum_{j_1=0}^\infty ... \sum_{j_I=0}^\infty
  \frac{\epsilon^{(j_1)}...\epsilon^{(j_I)}}{\theta^{(j_1+...+j_I)}}
  \frac{\beta^{j_1}}{j_1!} ... \frac{\beta^{j_I}}{j_I!}
  \\
  &=
  \sum_{j=0}^\infty
  \frac{\beta^j}{\theta^{(j)}}
  \sum_{j_1+...+j_I=j}
  \frac{\epsilon^{(j_1)}...\epsilon^{(j_I)}}{j_1!...j_I!}
  =\f11 (\gamma \theta; \theta; \beta),
\end{split}
\end{equation}
where $\gamma = I/K$ is the fraction of
nodes on the first (lower) plane, and in the last equality we used
\begin{equation}\label{pocchammer-as-binomial-sum}
  \sum_{j_1+...+j_I=j}
  \binom{j}{j_1 ... j_I}
  a_1^{(j_1)} ... a_I^{(j_1)}
  =(a_1+...+a_I)^{(j)},
\end{equation}
where the sum runs over all non-negative $j_i$ that sum up to $j$.

Now, consider a situation in which the first $i$ out of $k$ counts
happen to come from the first plane.
This means that they are among the first $I$ elements of $\vec\nu_\sigma$.
Since assigning these $i$ counts to different locations
within the first $I$ slots in the $\vec\nu_\sigma$ vector
(while keeping their original order)
does not change the result,
we can for convenience assign them to be the
first $i$ elements of $\vec\nu_\sigma$, followed by $I-i$ zeros.
Then
\begin{equation}
	\vec\nu_\sigma = (\underbrace{\overbrace{\nu_1, ..., \nu_i}^{i}, 0, ..., 0}_{I},
	\underbrace{\overbrace{\nu_{i+1}, ..., \nu_k}^{k-i}, 0, ..., 0}_{K-I}).
\end{equation}
The corresponding generalized confluent hypergeometric function is again
given by the sum over $I$ indices:
\begin{equation}
\begin{split}
  &\mathcal F(\vec\epsilon + \vec\nu_\sigma; |\vec\epsilon|+n; \vec\beta) \\
  &=
  \sum_{j_1=0}^\infty ... \sum_{j_I=0}^\infty
  \frac{\beta^{j_1}}{j_1!} ... \frac{\beta^{j_I}}{j_I!}
  \frac{(\epsilon+\nu_1)^{(j_1)} ... (\epsilon+\nu_i)^{(j_i)}
  \epsilon^{(j_{i+1})} ... \epsilon^{(j_I)}}
  {(K\epsilon + n)^{(j)}}.
\end{split}
\end{equation}
We can rewrite it as
\begin{equation}
\begin{split}
	\mathcal F(\vec\epsilon + \vec\nu_\sigma; |\vec\epsilon|+n; \vec\beta)
	&=
	\sum_{j=0}^\infty
	\frac{\beta^{j}}{j!}
	\sum_{j'+j''=j}
	\binom{j}{j'\; j''}
	\frac{1}{(K\epsilon + n)^{(j'+j'')}}
	\\
	&\times
	\sum_{j_1+...+j_i=j'}
	\binom{j'}{j_1 ... j_i}
	(\epsilon+\nu_1)^{(j_1)} ... (\epsilon+\nu_i)^{(j_i)}
	\\
	&\times
	\sum_{j_{i+1}+...+j_I=j''}
	\binom{j''}{j_{i+1} ... j_I}
	\epsilon^{(j_{i+1})} ... \epsilon^{(j_I)}.
\end{split}
\end{equation}
Using Eq.~\ref{pocchammer-as-binomial-sum} immediately leads to
\begin{equation} \label{mathcalF=F11:again}
	\mathcal F(\vec\epsilon + \vec\nu_\sigma; |\vec\epsilon|+n; \vec\beta)
	=
	\f11 \left(\gamma\theta + \sum_{m=1}^i \nu_m; \theta + n; \beta \right),
\end{equation}
so that the generalized confluent hypergeometric function is once again reduced to the
ordinary confluent hypergeometric function.

Lastly, we need to take into account the fact that we can put
counts into different positions of the $\vec\nu_\sigma$ vector.
This introduces an additional binomial pre-factor $\binom{I}{i}$.
Similarly, placing the rest of the counts into the last $K-I$ entries of the $\vec\nu_\sigma$ vector introduces another binomial pre-factor $\binom{K-I}{k-i}$.
Using these pre-factors together with Eqs.~\ref{mathcalF=F11} and \ref{mathcalF=F11:again} in Eq.~\ref{sampling-prob} yields
Eq.~\ref{sampling-prob-2-planes-finite-K}.

\section{Frequency spectrum for the arbitrary landscape} \label{appendix:frequency-spectrum}

We can expand the exponents in the allele frequency distribution
(Eq.~\ref{allele-frequency-distribution}) into a series:
\begin{equation}
  p(x_1, ..., x_K) =
  \frac{1}{B(\vec\epsilon) \mathcal F(\vec\epsilon; |\vec\epsilon|; \vec\beta)}
  \prod_{i=1}^K \sum_{j_i=0}^\infty
  \frac{\beta_i^{j_i}}{j_i !} x_i^{\epsilon+j_i-1},
\end{equation}
and apply
\begin{equation}\label{appendix:integral}
 \int_0^{1-s} x^{a-1} (1-s-x)^{b-1} dx
 =
 (1-s)^{a+b-1} \frac{\Gamma(a)\Gamma(b)}{\Gamma(a+b)}
\end{equation}
in order to get
\begin{equation}
\begin{split}
  p(x_{\sigma(1)}, ..., x_{\sigma(k)})
  =&
  \frac{\Gamma(|\vec\epsilon|)}
  {\Gamma(|\vec\epsilon_{\sigma^c}|) \prod_{i\in\sigma}\Gamma(\epsilon_i)}
  (1-|\vec x_\sigma|)^{|\vec \epsilon_{\sigma^c}|-1}
  \\
  \times&
  \frac{\mathcal F\left(\vec\epsilon_{\sigma^c};
  |\vec\epsilon_{\sigma^c}|;
  (1-|\vec x_\sigma|)\vec\beta_{\sigma^c}\right)}
  {\mathcal F(\vec\epsilon; |\vec\epsilon|; \vec\beta)}
  \prod_{i\in\sigma} x_i^{\epsilon_i-1} e^{\beta_i x_i},
\end{split}
\end{equation}
where $\sigma^c$ is a list of the $K-k$ alleles not contained in $\sigma$, 
and therefore $\vec\epsilon_{\sigma^c}$ and
$\vec\beta_{\sigma^c}$ are $(K-k)$-dimensional vectors
obtained from $\vec\epsilon$ and $\vec\beta$ by eliminating elements at the positions specified by $\sigma$, while
$\vec x_\sigma$ is a $k$-dimensional vector obtained from the $K$-dimensional vector $\vec x$ by keeping the elements at the positions specified by $\sigma$, and eliminating the rest.

With equal mutation rates, we have
\begin{equation}
\begin{split}
  p(x_{\sigma(1)}, ..., x_{\sigma(k)})
  &=
  \frac{\Gamma(K\epsilon)}{\Gamma((K-k)\epsilon)\Gamma(\epsilon)^k}
  (1-|\vec x_\sigma|)^{(K-k)\epsilon-1}
  \\
  &\times
  \frac{
  \mathcal F \left(\vec\epsilon_{\sigma^c};
  (K-k)\epsilon; (1-|\vec x_\sigma|)\vec\beta_{\sigma^c}\right)
  }
  {
  \mathcal F(\vec\epsilon; K\epsilon; \vec\beta)
  }
  \prod_{i\in\sigma}x_i^{\epsilon-1} e^{\beta_i x_i}.
\end{split}
\end{equation}
Taking the $k=1$ case and summing over allelic types, we obtain Eq.~\ref{frequency-spectrum-general}.

\bibliographystyle{nar}

\begin{thebibliography}{10}

\bibitem{Podgornaia2015}
Podgornaia, A. and Laub, M. (2015)
{\em Science} {\bf 347}, 673--677.

\bibitem{Lunzer2005}
Lunzer, M., Miller, S.~P., Felsheim, R., and Dean, A.~M. (2005)
{\em Science} {\bf 310}, 499--501.

\bibitem{Romero2009}
Romero, P.~A. and Arnold, F.~H. (2009)
{\em Nat Rev Mol Cell Biol} {\bf 10}, 866--876.

\bibitem{Lunzer2010}
Lunzer, M., Golding, G.~B., and Dean, A.~M. (2010)
{\em PLoS Genet} {\bf 6}, e1001162.

\bibitem{Breen2012}
Breen, M., Kemena, C., Vlasov, P., Notredame, C., and Kondrashov, F. (2012)
{\em Nature} {\bf 490}, 535--538.

\bibitem{Lind2010}
Lind, P.~A., Berg, O.~G., and Andersson, D.~I. (2010)
{\em Science} {\bf 330}, 825--827.

\bibitem{Hietpas2011}
Hietpas, R.~T., Jensen, J.~D., and Bolon, D. N.~A. (2011)
{\em Proc Nat Acad Sci USA} {\bf 108}, 7896--7901.

\bibitem{Sanjuan2004}
Sanjuan, R., Moya, A., and Elena, S.~F. (2004)
{\em Proc Nat Acad Sci USA} {\bf 101}, 8396--8401.

\bibitem{EyreWalker2007}
Eyre-Walker, A. and Keightley, P.~D. (2007)
{\em Nat Rev Genet} {\bf 8}, 610--618.

\bibitem{Wagner2008}
Wagner, A. (2008)
{\em Nat Rev Genet} {\bf 9}, 965--974.

\bibitem{ewens-04}
Ewens, W. (2004)
Mathematical Population Genetics: I. Theoretical Introduction,
Springer,  2nd edition.

\bibitem{ewens-72}
Ewens, W. (1972)
{\em Theor Pop Biol} {\bf 3}, 87--112.

\bibitem{Slatkin1994}
Slatkin, M. (1994)
{\em Genet Res Cambr} {\bf 64}, 71--74.

\bibitem{li-77}
Li, W.-H. (1978)
{\em Genetics} {\bf 90}, 349--382.

\bibitem{li-78}
Li, W.-H. (1977)
{\em Proc Nat Acad Sci USA} {\bf 74}, 2509--2513.

\bibitem{li-79}
Li, W.-H. (1979)
{\em Genetics} {\bf 92}, 647--667.

\bibitem{ewens-80}
Ewens, W. and Li, W.-H. (1980)
{\em J Math Biol} {\bf 10}, 155--166.

\bibitem{griffiths-83}
Griffiths, R. (1983)
{\em Journal of Mathematical Biology} {\bf 17}, 1--10.

\bibitem{ethier-87}
Ethier, S. and Kurtz, T. (1987)
{\em Stochastic Models in Biology, Lecture Notes in Biomathematics} {\bf 70},
  72--86.

\bibitem{joyce-95a}
Joyce, P. and Tavare, S. (1995)
{\em J Math Biol} {\bf 33}, 602--618.

\bibitem{joyce-95b}
Joyce, P. (1995)
{\em J Appl Prob} {\bf 32(3)}, 609--622.

\bibitem{grote-02}
Grote, M. and Speed, T. (2002)
{\em Ann Appl Prob} {\bf 12}, 637--663.

\bibitem{joyce-05}
Joyce, P., Genz, A., and Buzbas, E. (2012)
{\em J Comp Biol} {\bf 16(6)}, 650--661.

\bibitem{charlesworth-93}
Charlesworth, B., Morgan, M., and Charlesworth, D. (1993)
{\em Genetics} {\bf 134}, 1289--1303.

\bibitem{Hudson1994}
Hudson, R. and Kaplan, N. (1994)
Gene trees with background selection
In B. Golding, (ed.), Non-Neutral Evolution: Theories and Molecular Data,  pp.
  140--153 Chapman and Hall New York, NY.

\bibitem{desai-12}
Desai, M., Nicolaisen, L., Walczak, A., and Plotkin, J. (2012)
{\em Theor Pop Biol} {\bf 81}, 144--157.

\bibitem{sawyer-92}
Sawyer, S. and Hartl, D. (1992)
{\em Genetics} {\bf 132}, 1161--1176.

\bibitem{moran-58}
Moran, P. A.~P. (1958)
{\em Math Proc Cambr Philos Soc} {\bf 54}, 60--71.

\bibitem{Gillespie2004}
Gillespie, J. (2004)
Population Genetics: A Concise Guide,
The Johns Hopkins University Press, Baltimore.

\bibitem{wright-31}
Wright, S. (1931)
{\em Genetics} {\bf 16}, 97--159.

\bibitem{lassig-10}
Mustonen, V. and L\"assig, M. (2010)
{\em Proc Nat Acad Sci USA} {\bf 107(9)}, 4248--4243.

\bibitem{watterson-77}
Watterson, G. (1977)
{\em Genetics} {\bf 85}, 789--814.

\bibitem{kimura-62}
Kimura, M. (1962)
{\em Genetics} {\bf 47}, 713--719.

\bibitem{kimura-69}
Kimura, M. and Ohta, T. (1969)
{\em Genetics} {\bf 61}, 763--771.

\bibitem{Crow1970}
Crow, J. and Kimura, M. (1970)
An Introduction to Population Genetics Theory,
The Blackburn Press, Caldwell, NJ.

\bibitem{sella-05}
Sella, G. and Hirsh, A. (2005)
{\em Proc Nat Acad Sci USA} {\bf 102}, 9541--9546.

\bibitem{sella-09}
Sella, G. (2009)
{\em Theor Pop Biol} {\bf 75}, 30--34.

\bibitem{rouzine-01}
Rouzine, I.~M., Rodrigo, A., and Coffin, J.~M. (2001)
{\em Microbiol Mol Biol Rev} {\bf 65}, 151--185.

\bibitem{kingman-75}
Kingman, J. F.~C. (1975)
{\em Journal of the Royal Statistical Society, B} {\bf 37(1)}, 1--22.

\bibitem{kingman-77}
Kingman, J. F.~C. (1977)
{\em Theor Pop Biol} {\bf 11(2)}, 274--283.

\bibitem{etheridge-11}
Etheridge, A. (2011)
Some Mathematical Models from Population Genetics,
Springer-Verlag, Berlin 1st edition.

\bibitem{nielsen-05}
Nielsen, R. (2005)
{\em Annu Rev Genet} {\bf 39}, 197--218.

\bibitem{kullback-51}
Kullback, S. and Leibler, R. (1951)
{\em Ann Math Stat} {\bf 22}, 79--86.

\end{thebibliography}

\end{document}